\documentclass[twocolumn, prx, superscriptaddress, notitlepage]{revtex4-1}
\usepackage[utf8]{inputenc}
\usepackage{graphicx}
\usepackage{physics}
\usepackage{float} 
\usepackage{subfigure} 
\usepackage{color}
\usepackage{amsmath}
\usepackage{amsfonts}
\usepackage[normalem]{ulem}
\usepackage{xspace}
\usepackage[dvipsnames]{xcolor}
\usepackage{dcolumn}
\usepackage[breaklinks,colorlinks,linkcolor=blue,citecolor=blue,urlcolor=blue]{hyperref}
\usepackage{lineno}

\newcommand{\be}{\begin{equation}}
\newcommand{\ee}{\end{equation}}
\newcommand{\bea}{\begin{eqnarray}}
\newcommand{\eea}{\end{eqnarray}}

\begin{document}
\title{Excitonic skin effect}

\author{Wenhui Xu}

 \affiliation{Department of Physics and Astronomy, Purdue University, West Lafayette, IN, 47907, USA}

\author{Qi Zhou}
\email{zhou753@purdue.edu}
\affiliation{Department of Physics and Astronomy, Purdue University, West Lafayette, IN, 47907, USA}
\affiliation{Purdue Quantum Science and Engineering Institute, Purdue University, West Lafayette, IN 47907, USA}
\date{\today}

\begin{abstract}
%Skin effect is a drastic phenomena that imaginary vector potentials force an extensive number of physical states to concentrate near the boundaries of a system. Whereas conventional skin effect concerns single particles, it is desirable to investigate whether composite particles may also exhibit skin effect. Here, 
We show that strong interactions combined with band-dependent imaginary vector potentials give rise to boundary localization of particle-hole pairs, which we term the excitonic skin effect. In a bilayer system with layer-specific gain/loss and an in-plane magnetic field, excitons experience a net imaginary vector potential, resulting in directional amplification of particle-hole pairs. Including nearest-neighbor interactions leads to a non-Hermitian bosonic Kitaev model, where the pairing effects grow exponentially with the size of the system, revealing a unique form of critical skin effect in interacting systems.  Our framework applies to both atomic and electronic platforms and is directly testable in current experiments. These results also provide a route to explore non-Hermitian analogs of tensor gauge fields.

%and are thus anomalously localized near the boundaries, a phenomenon we dubbed exitonic skin effect. We show that dissipative spin-orbit coupling provides a natural platform to study such excitonic skin effect. Our results suggest an efficient means to harvest excitons and can also be generalized to produce multipolar skin effects. 

\end{abstract}
\maketitle

%\section{Introduction}
%Skin effect originally describes the localization of an alternating electric current near the surface of a conductor. 
The skin effect—a phenomenon where an extensive number of physical states accumulate at system boundaries—has emerged as a striking feature in both quantum and classical systems~\cite{yao2018,kunst2018,martinez2018,lee2019,borgnia2020,zhang2020,Okuma2020,Okuma2023,Lin2023}. In non-Hermitian systems,  this concentration is induced by non-reciprocal couplings that push eigenstates toward the boundaries~\cite{Nelson1993,Hatano1996,Hatano1997,Hatano1998}.
Such a non-Hermitian skin effect has enabled advances in quantum sensing, photonics and wave control~\cite{Regensburger2012,Feng2014,Wiersig2014,Longhi2015,Wiersig2016,Liu2016,Chen2017,Weidemann2020}. In parallel, it has been found that curved spaces naturally support skin effect~\cite{Zhang2021}. Because of the geometric distortion inherent in curved manifolds, wavefunctions tend to localize and amplify in certain regions. Skin effects also arise in classical systems. For example, random walks on tree graphs exhibit directional drift and boundary accumulation~\cite{Hughes1982, Cassi1990,Monthus1996}. 

Although these skin effects were initially studied in isolation, recent work has revealed a deep connection between them~\cite{Lv2022}. Central to this unification is the concept of an imaginary vector potential~\cite{Nelson1993,Hatano1996}.  In non-Hermitian systems, it emerges from non-reciprocal couplings; in curved spaces, it arises naturally from the Laplace-Beltrami operator~\cite{daCosta1982}; and in classical random walks, it manifests as a drift term. 
Unlike the ordinary real vector potentials, which imprint real phases to the wavefunctions, imaginary vector potentials induce position-dependent distortions, leading to the accumulation of physical states at boundaries, i.e., the skin effect. 

While the skin effect has been extensively studied in non-interacting systems, recent research has begun to explore its manifestations in correlated systems, where interactions enrich the non-Hermitian dynamics~\cite{Li2020,Yang2021,Shen2022,Lee2021,Kawabata2022,Alsallom2022}. Notably, the concept has been extended to multipolar excitations, such as dipoles and quadrupoles, via non-reciprocal correlated hopping~\cite{Gliozzi2024}. These developments pushes the study of skin effects into a new regime, where composite excitations exhibit boundary accumulation, opening possibilities for higher-order topological control and interaction-enabled non-Hermitian devices.
 
Here, we point out a new class of skin effect—the excitonic skin effect—arising from the interplay between interactions and band-dependent imaginary vector potentials. A prototypical system exhibiting this behavior is a bi-layer structure subject to an in-plane magnetic field and the layer-dependent loss or gain, which generates a band structure where the ground and excited bands carry imaginary vector potentials of equal amplitude but opposite direction. 
When strong onsite interactions bind a hole in the ground band with  a particle in the excited band to form an exciton, this 
composite experiences a finite net imaginary vector potential, leading to directional amplification and boundary accumulation of excitonic states—a hallmark of the excitonic skin effect. 

Including nearest neighbor interactions further enriches the physics. The system then supports pair creation and annihilation processes, which map onto a non-Hermitian bosonic Kitaev model in the hard-core boson limit. The imaginary vector potential amplifies the pairing terms, with their influence growing exponentially with the system size. This behavior can be viewed as an interacting counterpart to the critical skin effect.  In addition to electronic systems, our framework applies to spin-orbit coupled atoms under dissipation, making it also relevant for current ultracold atom experiments. It is also worth mentioning that viewing the energy as a synthetic dimension allows the band-dependent imaginary vector potential to be interpreted as a higher-rank imaginary tensor gauge field—an analog of real tensor gauge fields~\cite{Pretko2017,Ma2018,Bulmash2018,Zhang2025}. Our work thus opens up a new venue for studying the non-Hermitian analogs of tensor gauge fields in terms of the excitonic skin effect.

%We show that adding dissipation to spin-orbit couplings naturally provides such band-dependent imaginary vector potentials. This scheme can be straightforwardly generalized to produce skin effect for biexcitions. 

%\section{Results}
%To introduce the synthetic imaginary vector potential subject to excitons, 
\begin{figure}[tbp]
    \centering
    \includegraphics[width=1\columnwidth]{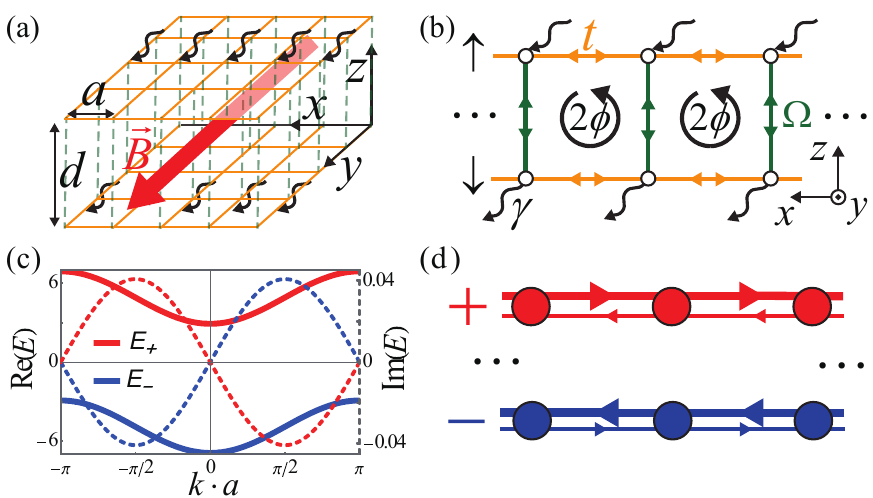}
    
    \caption{(a) A schematic of the bi-layer system with an in-plane magnetic field. (b) A two-leg ladder model describing the motion of particles in the $x$-$z$ plane. Wiggly arrows denote the gain and loss added to the top and lower layers. $2\phi$ is the magnetic flux per unit cell. (c) A typical single-particle dispersion. %of a single 2-level atom with SOC and dissipation. 
    Solid (dashed) curves denote $\Re{E_\pm}\sim k^2$ ($\Im{E_\pm}\sim \pm k$) near $k=0$. (d) The effective lattice model describes two decoupled Hatano-Nelson chains with opposite non-reciprocity.
    }
    \label{bilayer}
\end{figure}

We consider a bilayer system as shown in Fig.~\ref{bilayer}a. Two layers in the $x-y$ plane is separated from each other by a distance $d$ in the $z$-direction. A magnetic field is applied in the $y$-direction. Since the dynamics in the $y$-direction is regular and the skin effect exists only in the $x$-direction, we focus on the $x-z$ plane, which can be described by a two-leg ladder in Fig.~\ref{bilayer}b. We first start from the single-particle Hamiltonian, which is written as
%start from a lattice model,  %single-particle  Hamiltonian in a 1D toy model,
\begin{equation}
\begin{split}
   &\hat{H}_s=\hat{K}+\hat{H}_d,\\
   &\hat{K}=( \sum_{j,\sigma} -te^{i\phi_\sigma} \hat{c}^\dagger_{j,\sigma} \hat{c}_{j+1,\sigma} + \Omega \sum_j \hat{c}^\dagger_{j,\uparrow} \hat{c}_{j,\downarrow}+h.c.) , \\
   &\hat{H}_d=i\gamma\sum_j (\hat{c}^\dagger_{j,\uparrow} \hat{c}_{j,\uparrow}-\hat{c}^\dagger_{j,\downarrow} \hat{c}_{j,\downarrow}),
   %&\hat{V}=V_0\sum_{j}\hat{n}_{j,\uparrow}\hat{n}_{j,\downarrow}.  
   \end{split}\label{HL}
\end{equation}
where $\sigma=(\uparrow,\downarrow)$ is a pseudospin index that denotes the upper and lower layer, respectively. $\hat{c}^\dagger_{j,\sigma}$ ($\hat{c}_{j,\sigma}$) is the fermionic creation (annihilation) operator at site $j$ for spin-$\sigma$, and $\hat{n}_{j,\sigma}=\hat{c}^\dagger_{j,\sigma}\hat{c}_{j,\sigma}$.   %The first term represents nearest-neighbor tunneling with a spin-dependent phase, and the second term denotes onsite spin-flip. 
$\hat{K}$ is a two-leg Harper-Hofstadter model, $\phi_\uparrow=-\phi_{\downarrow}=\phi$ is the Peierls phase, $2\phi$ is the magnetic flux per plaquette, and $\Omega$ is the intra-lyaer tunneling strength. $\hat{H}_d$ denotes the dissipation in the system. The upper and lower layers have gain and loss rate, $\pm \gamma$, respectively. It is worth mentioning that the dissipative rates in these two layers do not have to be equal in magnitude and opposite in sign. We note that  $i\sum_\sigma\gamma_{\sigma}\hat{c}^\dagger_{j,\sigma} \hat{c}_{j,\sigma}=i\bar{\gamma}\sum_\sigma \hat{c}^\dagger_{j,\sigma}\hat{c}_{j,\sigma}+i\gamma(\hat{c}^\dagger_{j,\uparrow} \hat{c}_{j,\uparrow}-\hat{c}^\dagger_{j,\downarrow} \hat{c}_{j,\downarrow})$, where $\bar{\gamma}=(\gamma_\uparrow+\gamma_\downarrow)/2$ and $\gamma=(\gamma_\uparrow-\gamma_{\downarrow})/2$. Any finite $\gamma_{\uparrow}-\gamma_{\downarrow}$ thus leads to $\hat{H}_d$. The term proportional to $\bar{\gamma}$ only leads to a trivial overall decay of the total particle number and does not change our main results of the skin effect.

$\hat{H}_s$ in Eq.(\ref{HL}) is directly realizable in %both ultracold atomic systems and electronic systems in solids. In 
ultracold atoms, where $\hat{H}_s$ corresponds to the Hamiltonian of spin-orbit coupled fermions with dissipation, $\sigma=\uparrow,\downarrow$ represent two hyperfine spin states, and $\Omega$ denotes the Raman coupling strength. The phase carried by Raman lasers is imprinted onto atoms such that a two-leg Harper-Hofstadter ladder with a finite magnetic flux $2\phi$ is accessed~\cite{Lin2011,Galitski2013}. $\hat{H}_d$ with a finite $\gamma$ has recently been introduced in experiments~\cite{Ren2022,Tao2025}. In solids, $\hat{H}_s$ is a standard Hamiltonian describing two coupled layers with a parallel magnetic field when $\gamma=0$.
It applies to both ordinary bilayer systems or Moir\'e  superlattices. %In the latter case, the tunable $a$ of the superlattice provides experimentalists an extra tuning knob to control the ratio of the inter-layer and intra-layer interaction. 
A finite $\gamma$ can be accessed by introducing losses to one layer through photon or electric field induced emission of electrons. Alternatively, layer-dependent optical driving leading to excitations to higher energy bands could also effectively result in the desired losses. 

%layer-dependent gains or losses~\cite{x}.  

%When $\gamma=0$, this is a two-leg Harper-Hofstadter model and $\phi_\sigma$ determines the magnetic flux per plaquette~[XX].  A finite $\gamma$ corresponds to spin-dependent losses or gains, which have been implemented in multiple experiments~[XX]. 

%We first consider the single-particle part of the Hamiltonian $\hat{K}+\hat{H}_d$. 

A Fourier transform of $\hat{H}_s$ provides the Hamiltonian in the momentum space, 
\begin{equation}
   \hat{h}_k=\left(
\begin{array}{cc}
    -2t\cos [(k-k_0)a]+i\gamma &  \Omega  \\
    \Omega & -2t\cos[(k+k_0)a]    -i\gamma \\
\end{array}\right),\label{Hk}
\end{equation} 
where $k$ is the crystal momentum, $a$ is the lattice spacing in the $x$-direction, and $k_0=-\phi/a$ denotes the shift of the crystal momentum when the magnetic field is implemented. %$\pm i\gamma$ denotes spin-dependent gain/loss, and $\Omega$ is the Raman coupling strength. 
$\hat{h}_k$ can be easily diagonalized and 
the two eigenenergy bands $\{| +\rangle,|-\rangle \}$ have eigenenergies $E_{\pm}(k)$. When $\gamma\neq 0$, $E_{\pm}(k)$ becomes complex. A profound feature of $E_{\pm}(k)$ is that ${\text Re} E_{\pm}(k)\sim k^2$, ${\text Im} E_{\pm}(k)\sim k$ in a broad parameter regime. For instance, when $|\Omega|\gg|\gamma|$, $|\Omega|\gg|t|$ and $|ka|\ll1$ are satisfied, $E_\pm(k)$ has a simple analytical form, 
\begin{equation}
   E_\pm(k)=\alpha(k- i A_\pm)^2+O(k^3a^3)+C_\pm \label{bandve},
\end{equation}
where $\alpha=a^2t\cos (k_0a)$, $A_\pm= \frac{\pm \gamma  \tan (k_0 a) }{a \Omega}$, and $C_\pm$ are momentum-independent constants.
%are well-approximated by 
%\begin{equation}
%   E_\pm(k)=\alpha_\pm(k- i A_\pm)^2+O(k^3a^3)+C_\pm \label{bandve},
%\end{equation}
%where $\alpha_\pm=a^2t(\cos k_0a\pm\frac{2\Omega^2t\sin^2k_0a}{(\Omega^2-\gamma^2)^{3/2}})$, $A_\pm= \frac{\pm a \gamma t \sin k_0 a }{\alpha_\pm \sqrt{\Omega^2-\gamma^2}}$, and $C_\pm$ are momentum-independent constants .  
A notable feature of $E_\pm(k)$ is that $A_\pm$ arise in the energy as imaginary vector potentials. Unlike a real vector potential that shifts the band structure in the real momentum axis, here, an imaginary vector potential provides a linear imaginary part of the energy and the real part remains quadratic, as shown in Fig.~\ref{bilayer}c.

A previous study has considered the imaginary vector potential in a single band and the subsequent single-particle skin effects~\cite{Zhou2022}. Here, we focus on a unique property of Eq.(\ref{bandve}) that the ground and the excited bands are equipped with imaginary vector potentials of the same amplitudes and opposite directions. The origin of such a band-dependent imaginary vector potential can be unfolded if we rewrite Eq.(\ref{HL}) in a different basis. We define $\hat{c}_{j,\pm}=( \hat{c}_{j,\uparrow}\pm \hat{c}_{j,\downarrow})/\sqrt{2}$. In the limit $|\Omega|\gg |\gamma|$ and $|\Omega|\gg |t|$, a standard perturbation approach gives rise to an effective Hamiltonian, 
\begin{equation}
   \hat{K}_{e}= \sum_{j,s=\pm} -(t_{R,s}\hat{c}_{j+1,s}^\dagger \hat{c}_{j,s}+t_{L,s}\hat{c}_{j,s}^\dagger \hat{c}_{j+1,s})+\Omega (\hat{n}_{+}-\hat{n}_{-}), \label{Ke}
\end{equation}
where $t_{R,+}=t_{L,-}=t(\cos\phi+\gamma \sin\phi/\Omega)$, $t_{L,+}=t_{R,-}=t (\cos\phi-\gamma \sin\phi/\Omega)$, and $\hat{n}_{s}=\sum_j \hat{c}_{j,s}^\dagger \hat{c}_{j,s}$.
The effective Hamiltonian corresponds to two decoupled Hatano-Nelson chains with non-reciprocal tunneling $t_{L,s}\neq t_{R,s}$, as shown in Fig.~\ref{bilayer}d. %When $\phi\rightarrow0$, the tunneling strength is related with $A_{\pm}$ in Eq.(\ref{bandve}) by $t_{R,\pm}/t_{L,\pm}=e^{-2A_{\pm}a}$.
Such non-reciprocal tunnelings correspond to imaginary vector potentials $A_\pm$, which can be extracted from $t_{R,\pm}=\bar{t} e^{-A_{\pm}a}$ and $t_{L,\pm}=\bar{t} e^{A_{\pm}a}$. Here $\bar{t}=\sqrt{\cos^2\phi-\gamma^2\sin^2\phi/\Omega^2}t$, and 
\begin{equation}
 A_{\pm}=\frac{1}{2a} \ln({t_{L,\pm}/t_{R,\pm}})=\frac{\mp \gamma  \tan \phi }{a \Omega}+O(\gamma/\Omega)^3,\label{Apm}
\end{equation}
consistent with the result in the continuum limit in Eq.(\ref{bandve}). $A_\pm$ give rise to the well-celebrated non-Hermitian skin effect. The two chains have an energy difference $2\Omega$, corresponding to the ground and excited bands. Since the imaginary vector potentials of these two bands point towards opposite directions, a band-dependent skin effect arises, and eigenstates in the ground and excited bands are localized at opposite edges of the systems. 

%Whereas skin-effect in non-interacting systems has been well studied, here, we add onsite interactions to the Hamiltonian, 
%\begin{equation}  H_L= H^0_L+U\sum_n n_{\uparrow}n_{\downarrow}\label{HLint},
%\end{equation} where $n_{\sigma}=c_{n,\sigma}^\dagger c_{n,\sigma}$. The effective Hamiltonian $H_{eff}$ changes correspondingly, 

We now take into account the interaction $\hat{H}_{int}$. 
While atoms, in general, have short-range interactions, some magnetic species or the Rydberg states exhibit dipole-dipole interactions, which may lead to considerable nearest-neighbor interactions~\cite{Griesmaier2005,Stuhler2005,Gallagher2008,Browaeys2016}. As for electrons, the generic expression for the Coulomb interaction $\hat{H}_{col}=\hat{V}+\hat{U}$ is written as $\hat{V}=\tilde{V}_0\sum_{j,m}(m^2+d^2/a^2)^{-1/2}\hat{n}_{j+m,\uparrow}\hat{n}_{j,\downarrow}$, and $\hat{U}=\tilde{U}_0\sum_{j,m\neq0}|m|^{-1}(\hat{n}_{j+m,\uparrow}\hat{n}_{j,\uparrow}+\hat{n}_{j+m,\downarrow}\hat{n}_{j,\downarrow})$. 
$\hat{U}$ is the intra-layer Coulomb interaction that decays as $1/m$ as the separation between two sites $m$ increases, and $\hat{V}$ is the inter-layer Coulomb interaction. In Moir\'e  systems, the tunable lattice spacing of the superlattice provides experimentalists with an extra tuning knob to control $a/d$, the ratio of the inter-layer and intra-layer interaction. As such, without loss of generality, we consider the interaction $\hat{H}_{int}=\hat{V}_{on}+\hat{V}_{nn}+\hat{U}_{nn}$,
\begin{equation}
 \begin{split}   
  & \hat{V}_{on}= V_0\sum_j \hat{n}_{j,\uparrow}\hat{n}_{j,\downarrow},\\ 
&\hat{V}_{nn}=V_1\sum_{j}\hat{n}_{j\pm 1,\uparrow}\hat{n}_{j,\downarrow},\\
  & \hat{U}_{nn}=U_0\sum_{j}(\hat{n}_{j\pm1 ,\uparrow}\hat{n}_{j,\uparrow}+\hat{n}_{j\pm 1,\downarrow}\hat{n}_{j,\downarrow}), 
\end{split}
   \end{equation}
where $V_0=\tilde{V}_0{a}/{d}$, $V_1=\tilde{V}_0(1+d^2/a^2)^{-1/2}$ and $U_0=\tilde{U}_0$. 

\begin{figure}[tbp]
    \centering
    \includegraphics[width=1\columnwidth]{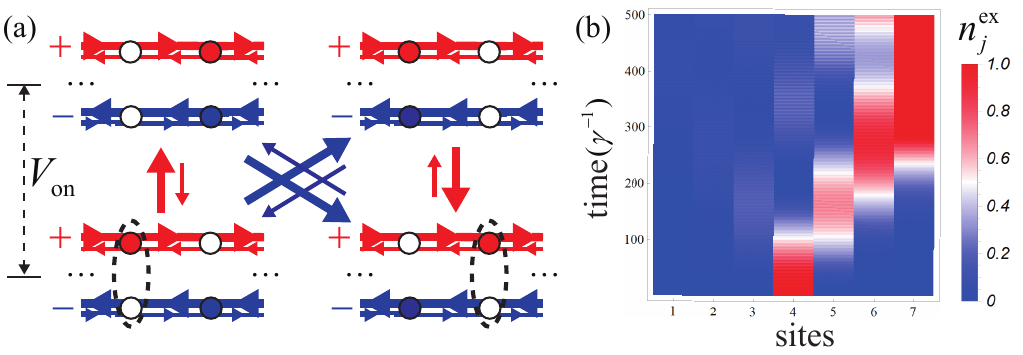}

    \caption{{\bf } {(a) %A single exciton is prepared on a fully filled ground band. 
    The non-reciprocal tunneling of a single exciton via a second-order process. (b) The exciton density $n_j^{ex}$ as a function of time and position. The result comes from the simulation of a $7\times 2$ full ladder model $\hat{H}_{s}+\hat{V}_{on}$, where we chose $\gamma=1$, $\Omega=10, V_0=100, t=1$, and $\phi=1$.}}
    \label{dn}
\end{figure}

Since the interaction decays as the separation between two sites increases, we first consider the dominant term, the onsite inter-layer interaction,  which is rewritten as%in terms of $\hat{n}_{j,s}=\hat{c}^\dagger_{j,s}\hat{c}_{j,s}$,
\begin{equation}
    \hat{V}_{on}= V_0 \sum_j \hat{n}_{j,+}\hat{n}_{j,-}.\label{Von}
\end{equation}
%\begin{equation}
%\begin{split}
%   &\hat{U}=U_0\sum_{n,m\neq 0}{|2m|}^{-1} \{(\hat{n}_{n,+}+\hat{n}_{n,-})(\hat{n}_{n+m,+}+\hat{n}_{n+m,-})\\
%   &+(d_n^\dagger+d_n)(d_{n+m}^\dagger+d_{n+m})\},\\
%   &\hat{V}g=V_0\sum_{n,m}\frac{1}{4\sqrt{m^2+({d}/{a})^2}}\{(\hat{n}_{n,+}+\hat{n}_{n,-})(\hat{n}_{n+m,+}+\\
%   &\hat{n}_{n+m,-})-(d_n^\dagger+d_n)(d_{n+m}^\dagger+d_{n+m})\}.  
%   \end{split}\label{H_L}
%\end{equation}
%We thus see that the form of interactions remain unchanged. %particles in the ground and excited bands interact with each other by an onsite interaction of an amplitude $U$. 
The form for $\hat{V}_{on}$ remains unchanged as we rewrite it in terms of $\hat{n}_{j,\pm}=\hat{c}^\dagger_{j,\pm}\hat{c}_{j,\pm}$.
We consider a particle-hole excitation on a fully filled ground band. In the limit of $|V_0 |\gg |t|$, the particle in the excited band and the hole left in the ground band prefer to stay at the same lattice site to avoid the extra energy cost for a particle-hole pair to stay at different lattice sites. In other words,  an exciton is formed. This exciton can tunnel in the lattice via a second-order process, as shown in Fig.~\ref{dn}a. The effective Hamiltonian for a single exciton is written as,
\begin{equation}
    \hat{H}_{ex}=\sum_j J_R\hat{d}_{j+1}^\dagger \hat{d}_j+J_L\hat{d}_{j-1}^\dagger \hat{d}_j,\label{Heff}
\end{equation}
where $\hat{d}_j=\hat{c}_{j,-}^\dagger \hat{c}_{j,+}$ and $J_{R(L)}=\frac{-2t^2}{V_0}(\cos\phi\pm\gamma\sin\phi/\Omega)^2$. We see that the tunneling of an exciton is also non-reciprocal since $J_{R}\neq J_L$. Such nonreciprocal tunneling implies that an exciton is subject to a finite imaginary vector potential $A_{ex}$. From $J_{R(L)}=\bar{J}e^{\mp A_{ex}a}$, we obtain
\begin{equation}
  A_{ex}=\frac{1}{2a} \ln({J_{L,\pm}/J_{R,\pm}}).  
\end{equation} 
A straightforward calculation shows that $A_{ex}=A_{+}-A_{-}$,
where $A_\pm$ is the band-dependent imaginary vector potential of single particles in Eq.~(\ref{Apm}). This can be understood from the fact that a hole in the ground band is subject to an imaginary vector potential of the opposite direction compared to a particle in the same band, i.e., $-A_{-}$. The net imaginary vector potential applied to an exciton is thus the difference between $A_+$ and $A_-$.

%It is desirable to work out the effective Hamiltonian for excitons.
%a particle and a hole prefer to stay in the same lattice site, and hop together in the lattice to avoid the extra energy cost when a particle and a hole stay at different lattice site. %(Here we emphasize that when $|V_0|$ is comparable with $|\Omega|$, the system will experience several exceptional points and therefore has complex eigenvalues in certain parameter regimes. These interaction-induced EPs originate from the non-reciprocal tunnelings $t_{R,s}$ and $t_{L,s}$ that couple the eigenstates seperated by energy gaps $\sim V_0$. The competition between $\Omega$ and $V_0$ allows the presence of complex eigenvalues even with a small $t$.) 

%As we obtained the effective Hamiltonian in Eq.~(\ref{Heff}), 
A finite imaginary vector potential $A_{ex}$ acting on an exciton is expected to induce the excitonic skin effect. To confirm this, we numerically simulate the dynamics governed by the full model $\hat{H}_s+\hat{V}_{on}$. We consider the initial state 
\begin{equation}
    |\psi(0)\rangle =\hat{c}_{j_0,+}^\dagger \hat{c}_{j_0,-}|v\rangle,\label{init}
\end{equation}
where $|v\rangle=\prod_j \hat{c}_{j,-}^\dagger|0\rangle$ is an insulating state with the ground band fully occupied. $|\psi(0)\rangle $ thus corresponds to an excitation of a single exciton that occupies the $j_0$-th site. We numerically computed the time evolution of this initial state,
$|\psi(t)\rangle=e^{-i(\hat{H}_s+\hat{V}_{on})t}|\psi(0)\rangle$,
%this many-body problem 
on a $7\times 2$ ladder where it has 7 sites along the x-direction, and 2 sites along the y-direction representing the ground and excited band. The exciton is prepared in the middle of the ladder at $j_0=4$.
Using exact diagonalization, we %examine the time evolution of the 
obtain the time dependence of the
exciton density  $n^{ex}_j(t)\equiv\langle\psi(t)|\hat{d}_j^\dagger \hat{d}_j|\psi(t)\rangle$ in Fig.~\ref{dn}b. When $t=0$, $n^{ex}_{j}=\delta_{j,j_0}$ . As time goes by, the exciton propagates in the lattice. A notable feature is that this propagation is not symmetric and gets amplified toward the right, and therefore the distribution of the exciton density  $n^{ex}_j(t)$ has larger probability on the right-hand side. 
This is the characteristic feature of the skin effect in the presence of an imaginary vector potential $A_{ex}<0$ acting on the exciton. Unlike the boundary accumulation of single-particles in the ordinary skin effect, here, particle-hole pairs concentrate at the edge, signifying a new type of skin effect for excitons. %until it reaches the boundary and begins to bounce back at $\gamma t\approx 420$. This asymmetry of the exciton density supports the excitonic skin effect. 
It is worth pointing out that, 
%Despite that the density of excitons is amplified 
in spite of the directional amplification, %toward one direction, 
conserved quantities exist once we appropriately take into account the finite curvatures underlying Hatano-Nelson chains (supplementary materials).

Whereas the above discussions have clearly demonstrated the excitonic skin effect, it is useful to consider such results from a different perspective. In the language of synthetic dimensions, the energy could be regarded as an extra dimension $w$ perpendicular to the real dimensions. Eq.~(\ref{Ke}) thus includes an imaginary rank-2 tensor gauge field, $A_{wx}=\partial_w A_x=A_+-A_-$. Recent studies of fracton phases of matter~\cite{Yoshida2013,Vijay2015,Vijay2016,Pretko2017,Ma2018,Prem2018,Bulmash2018,Nandkishore2019,Pretko2020,Yuan2020,Lake2022} have shown that, when single-particle excitations become immobile, the couplings between higher-rank gauge fields with dipoles and other multipoles become critical. However, those studies have mainly focused on real tensor gauge fields. Here, an imaginary tensor gauge field $A_{wx}$ provides excitons with a complex Peierls phase, leading to an effective non-Hermitian ring exchange interaction, 
\begin{equation}
\begin{aligned}
    \hat{H}_{ring}\sim&\sum_j e^{-A_{wx}a}\hat{c}_{j+1,+}^\dagger \hat{c}_{j+1,-} \hat{c}_{j,-}^\dagger \hat{c}_{j,+} \\
    &+e^{A_{wx}a}\hat{c}_{j,+}^\dagger \hat{c}_{j,-} \hat{c}_{j+1,-}^\dagger \hat{c}_{j+1,+}.  
\end{aligned} 
 \end{equation}
 %which is identical to $\hat{H}_{ex}$.
The excitonic skin effect thus could also be regarded as a consequence of an imaginary rank-2 tensor gauge field. 

Now we consider effects beyond the on-site interaction. To this end, %a generic Coulomb interaction, \begin{equation} \begin{aligned} \hat{U}=&U_0\sum_{j,m\neq0}|m|^{-1}(\hat{n}_{j+m,\uparrow}\hat{n}_{j,\uparrow}+\hat{n}_{j+m,\downarrow}\hat{n}_{j,\downarrow}) \\\hat{V}=&V_0\sum_{j,m}(m^2+d^2/a^2)^{-1/2}\hat{n}_{j+m,\uparrow}\hat{n}_{j,\downarrow}.\end{aligned}\end{equation} Here $\hat{U}$ is the intra-layer Coulomb interaction that decays as $1/m$ as the separation between two sites $m$ increases, $\hat{V}$ is the inter-layer Coulomb interaction, $d$ is the distance between two layers and $a$ is the distance between nearest neighbor sites in the same layer.When expressed 
we express $\hat{H}_{int}$
in terms of $\hat{c}_{j,\pm}$ and $\hat{c}_{j,\pm}^\dagger$, then the nearest-neighbor interactions can be written as
\begin{equation}
\begin{split}
   \hat{U}_{nn}=&U_0\sum_{j} \{(\hat{n}_{j,+}+\hat{n}_{j,-})(\hat{n}_{j+1,+}+\hat{n}_{j+1,-})+\\
   &(\hat{d}_j^\dagger+\hat{d}_j)(\hat{d}_{j+1}^\dagger+\hat{d}_{j+1})\},\\
   \hat{V}_{nn}=&\frac{V_1}{2}\sum_{j}\{(\hat{n}_{j,+}+\hat{n}_{j,-})(\hat{n}_{j+1,+}+\hat{n}_{j+1,-})-\\
   &(\hat{d}_j^\dagger+\hat{d}_j)(\hat{d}_{j+1}^\dagger+\hat{d}_{j+1})\}.
   \end{split}
\end{equation}
%\begin{equation}
%\begin{split}
%   &\hat{U}=U_0\sum_{j,m\neq 0}{|2m|}^{-1} \{(\hat{n}_{j,+}+\hat{n}_{j,-})(\hat{n}_{j+m,+}+\hat{n}_{j+m,-})\\
%   &+(\hat{d}_j^\dagger+\hat{d}_j)(\hat{d}_{j+m}^\dagger+\hat{d}_{j+m})\},\\
 %  &\hat{V}=V_0\sum_{j,m}\frac{1}{4\sqrt{m^2+({d}/{a})^2}}\{(\hat{n}_{j,+}+\hat{n}_{j,-})(\hat{n}_{j+m,+}+\\
  % &\hat{n}_{j+m,-})-(\hat{d}_j^\dagger+\hat{d}_j)(\hat{d}_{j+m}^\dagger+\hat{d}_{j+m})\}.
 %  \end{split}
%\end{equation}
Consider the constraint that the total number of fermions per site is fixed at 1,  $\hat{n}_{j,+}+\hat{n}_{j,-}$ is equivalent to an identity operator for any $j$. The first terms in the expressions for $\hat{U}_{nn}$ and $\hat{V}_{nn}$ can then be dropped off. We thus only need to concentrate on the terms that depend on the dipole operators $\hat{d}_j$ and $\hat{d}_j^\dagger$. These terms contain hopping, pair creation and annihilation of excitons on nearest-neighbor sites. The effective Hamiltonian for excitons is written as
\begin{equation}
    \begin{aligned}
        \tilde{H}_{ex}=&\sum_j-\tilde{J}_R \hat{d}_{j+1}^\dagger \hat{d}_j-\tilde{J}_L \hat{d}_j^\dagger \hat{d}_{j+1}+\mu \hat{d}_j^\dagger \hat{d}_j\\
        &+\Delta(\hat{d}_j^\dagger \hat{d}_{j+1}^\dagger+\hat{d}_{j+1}\hat{d}_j),
    \end{aligned}\label{Hex}
\end{equation}
where $\Delta=U_0-V_1/2$, $\mu=2\Omega$, and $\tilde{J}_{R(L)}=\frac{2t^2 d}{V_0 a}(\cos \phi \pm\frac{\gamma \sin \phi}{\Omega})^2-\Delta$. %Eq.~(\ref{Hex}) may bring us even more interesting phenomena that Eq.~(\ref{Heff}) does not exhibit. 
When $\tilde{J}_L=\tilde{J}_R$, the above equation has the same form as the bosonic Kitaev model~\cite{McDonald2018}. Here, $\tilde{J}_L\neq \tilde{J}_R$, and each exciton should be regarded as a hard-core boson. $\tilde{H}_{ex}$ thus corresponds to a non-Hermitian bosonic Kitaev model in the hard-core limit. Alternatively, by mapping the hard-core bosons to spin-1/2s, Eq.~(\ref{Hex}) could be regarded as a non-Hermitian generalization of the spin Kitaev model. 

\begin{figure}[tbp]
    \centering
    \includegraphics[width=1\columnwidth]{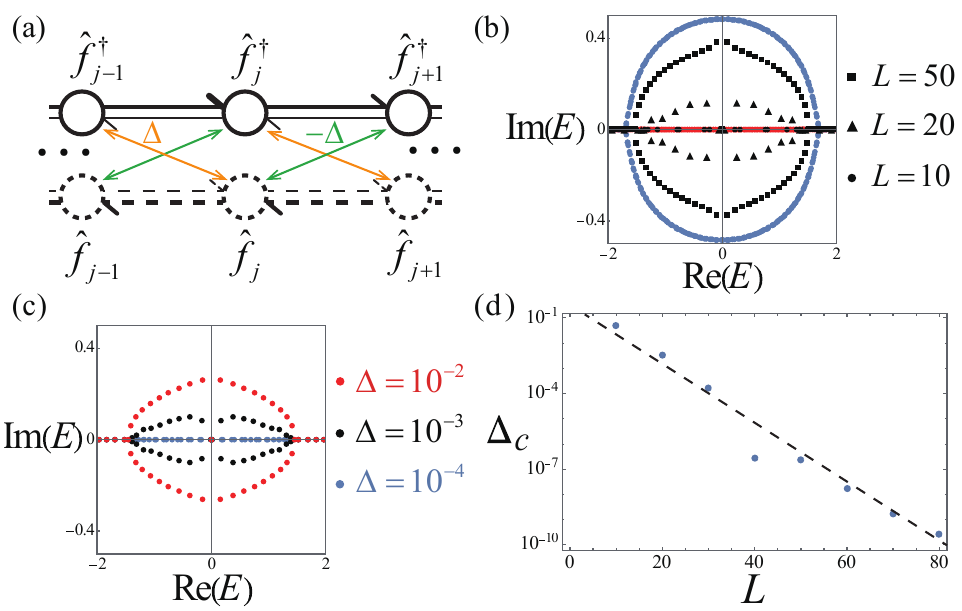}

    \caption{{(a) %Two Hatano-Nelson chains with opposite non-reciprocity experience the diagonal couplings $\Delta$ between nearest-neighbor sites of these two chains. 
    The pairing between nearest-neighbor sites leads to two coupled Hatano-Nelson chains, one for particles and the other for holes.  (b) The OBC eigenenergy spectra of $\hat{H}_f$ with system size $L=10$, 20, and 50. Here we choose $\tilde{J}_L=0.75$, $\tilde{J}_R=1.25$, $\mu=0.5$ and $\Delta=0.01$. Blue dots denote the results in the limit $L\rightarrow \infty$. For comparison, Red dots show the results with a vanishing $\Delta$ when  $L\rightarrow \infty$. (c) The OBC spectra of $\hat{H}_f$ with coupling strength $\Delta=10^{-2}$, $10^{-3}$, and $10^{-4}$. Here we fix $L=30$.   (d) The scaling relation between the critical coupling $\Delta_c$ and the system size $L$.}}
    \label{CNHSE}
\end{figure}

A unique feature distinguishing $\tilde{H}_{ex}$ from its Hermitian counterparts is that the non-reciprocal tunneling significantly amplifies the pairing term. To show this drastic effect transparently, we solve the energy spectrum of $\tilde{H}_{ex}$ by
mapping it to a fermionic model $\hat{H}_f$, which has the same energy spectrum as $\hat{H}_{ex}$, via a Jordan-Wigner transformation $\hat{d}_j^\dagger=\exp(-i\pi \sum_{n<j}\hat{f}_n^\dagger \hat{f}_n)\hat{f}_j^\dagger$. The last two terms in Eq.({\ref{Hex}) become the $p$-wave pairing for fermions, and $\hat{H}_f$ is a non-Hermitian generalization of the 1D Kitaev model~\cite{Kitaev2001}. The conditions for the existence of Majorana zero modes %and non-Hermitian skin effects in the Kitaev model with non-reciprocal tunnelings 
in such a model have been previously discussed~\cite{Li2022,Yan2023,Ardonne2025}. Here, we focus on how the non-reciprocal coupling and the pairing term control the energy spectrum. 

%On the other hand, 
$\hat{H}_f$ in the Nambu space is written as
\begin{equation}
    \hat{H}_f=\frac{1}{2}\hat{\Psi}_f^\dagger\left( \begin{array}{cc}
        \hat{H}_{11} & \hat{H}_{12} \\
        \hat{H}_{12}^T & -\hat{H}_{11}^T
    \end{array}\right)\hat{\Psi}_f,\label{Hf}
\end{equation} where  $\hat{H}_{11}=\text{Trid} [\mu, -J_L, -J_R]$,  $\hat{H}_{12}=\text{Trid} [0, \Delta, -\Delta]$, and $\hat{\Psi}_f=(\hat{f}_1,\cdots,\hat{f}_L,\hat{f}_1^\dagger,\cdots,\hat{f}_L^\dagger)^T$. ${\text{Trid} [a, b, c]}$ is a triadiagonal matrix,
%\begin{equation} \hat{H}_{11}=\left(\begin{array}{cccc}   \mu & -J_L & &\\      -J_R & \ddots & \ddots &\\        & \ddots & \ddots & -J_L\\      & & -J_R & \mu  \end{array}\right),
%\end{equation}
%\begin{equation}   \hat{H}_{12}=\left(\begin{array}{cccc}         0 & \Delta & &\\ -\Delta & \ddots & \ddots &\\        & \ddots & \ddots & \Delta\\     & & -\Delta & 0 \end{array}\right), \end{equation}
\begin{equation}
 {\text{Trid} [a, b, c]}=\left(\begin{array}{cccc}
           a & b & &\\
           c & \ddots & \ddots &\\
            & \ddots & \ddots & b\\
            & & c & a
       \end{array}\right).
\end{equation}
 We note that the two diagonal blocks can be interpreted as two Hatano-Nelson chains with opposite non-reciprocity, one for the particle and the other for the hole in the $p$-wave superconductor. As such, the pairing term 
$\Delta$ can be viewed as the diagonal couplings between nearest-neighbor sites of these two chains, as shown in Fig.~\ref{CNHSE}a. Since this is similar to the critical non-Hermitian skin effect, where onsite couplings exist between two Hatano-Nelson chains with opposite non-reciprocity~\cite{Li2020}, it is desirable to explore whether the critical skin effect may exist in this interacting system. 

We numerically solve the energy spectrum of $\hat{H}_f$ with the open boundary condition (OBC).
As shown in Fig.~\ref{CNHSE}c, for a finite system with a fixed number of sites $L$, when $\Delta=0$, all eigenenergies are real. Increasing $\Delta$ to a critical value $\Delta_c$, some eigenenergies become complex. In particular, the larger $L$ is, the easier it is for a system with a fixed $\Delta$ to acquire complex eigenenergies, as shown in Fig.~\ref{CNHSE}b. In other words, $\Delta_c$ decreases exponentially with increasing $L$.
%$\Delta$ can be exponentially enhanced by the system size $L$. As shown in Fig~\ref{CNHSE}, the eigenenergy spectra for two decoupled chains($\Delta=0$, red) and coupled chains($\Delta=0.01$, blue) are qualitatively different in the thermodynamic limit. For a finite system size $L$, the physical eigenenergy spectrum (black) interpolates between the coupled and decoupled OBC spectra. There exists a critical coupling strength $\Delta_c(L)$, such that all the eigenenergies are real at $\Delta\leq\Delta_c$, while complex eigenenergy emerges at $\Delta>\Delta_c$. 
The scaling behavior of $\Delta_c$ is shown in Fig.~\ref{CNHSE}d, indicating an exponential enhancement of the pairing $\Delta$ as $L$ increases. 
We thus conclude that %the nearest-neighbor intra-layer interaction or 
the nearest-neighbor interaction and the non-reciprocal tunneling %nter-layer interaction can 
give rise to an interacting counterpart of the critical non-Hermitian skin effect.

We have shown that the excitonic skin effect arises from the interplay of magnetic fields, interactions, and dissipation. This phenomenon is potentially observable in both atomic and electronic systems. Exploring this new type of skin effect may expand the frontier in the study of higher-rank tensor gauge fields to a complex domain. It will also provide a new platform for physicists to study mon-Hermitian physics of composite particles

%$E_{\pm}(k)=(k^2+k_0^2)/2\pm\sqrt{k_0^2k^2-2i\gamma k_0 k-\gamma^2+\Omega^2}$,

%In certain parameter regimes, the two dressed energy bands $\{| +\rangle,|-\rangle \}$ can be approximated by a quadratic form near the band bottom. To be more specific, when %|\Omega|>|\gamma|$ and $\epsilon\equiv\frac{|k_0 k|}{\sqrt{\Omega^2-\gamma^2} }\ll1$, 

%whose eigenvalues are $E_{\pm}(k)=(k^2+k_0^2)/2\pm\sqrt{k_0^2k^2-2i\gamma k_0 k-\gamma^2+\Omega^2}$,

The authors thank Ian Spielman, Emmanuel Gutierrez and Yihang Zeng for helpful discussions. This work is supported by The U.S. Department of Energy, Office of Science through the Quantum Science Center (QSC), a National Quantum Information Science Research Center, and the Air Force Office of Scientific Research under award number FA9550-23-1-0491.

\bibliographystyle{apstest}
\bibliography{dc.bib}

\begin{thebibliography}{60}%
\makeatletter
\providecommand \@ifxundefined [1]{%
 \@ifx{#1\undefined}
}%
\providecommand \@ifnum [1]{%
 \ifnum #1\expandafter \@firstoftwo
 \else \expandafter \@secondoftwo
 \fi
}%
\providecommand \@ifx [1]{%
 \ifx #1\expandafter \@firstoftwo
 \else \expandafter \@secondoftwo
 \fi
}%
\providecommand \natexlab [1]{#1}%
\providecommand \enquote  [1]{#1}%
\providecommand \bibnamefont  [1]{#1}%
\providecommand \bibfnamefont [1]{#1}%
\providecommand \citenamefont [1]{#1}%
\providecommand \href@noop [0]{\@secondoftwo}%
\providecommand \href [0]{\begingroup \@sanitize@url \@href}%
\providecommand \@href[1]{\@@startlink{#1}\@@href}%
\providecommand \@@href[1]{\endgroup#1\@@endlink}%
\providecommand \@sanitize@url [0]{\catcode `\\12\catcode `\$12\catcode `\&12\catcode `\#12\catcode `\^12\catcode `\_12\catcode `\%12\relax}%
\providecommand \@@startlink[1]{}%
\providecommand \@@endlink[0]{}%
\providecommand \url  [0]{\begingroup\@sanitize@url \@url }%
\providecommand \@url [1]{\endgroup\@href {#1}{\urlprefix }}%
\providecommand \urlprefix  [0]{URL }%
\providecommand \Eprint [0]{\href }%
\providecommand \doibase [0]{https://dx.doi.org}%
\providecommand \selectlanguage [0]{\@gobble}%
\providecommand \bibinfo  [0]{\@secondoftwo}%
\providecommand \bibfield  [0]{\@secondoftwo}%
\providecommand \translation [1]{[#1]}%
\providecommand \BibitemOpen [0]{}%
\providecommand \bibitemStop [0]{}%
\providecommand \bibitemNoStop [0]{.\EOS\space}%
\providecommand \EOS [0]{\spacefactor3000\relax}%
\providecommand \BibitemShut  [1]{\csname bibitem#1\endcsname}%
\let\auto@bib@innerbib\@empty
%</preamble>
\bibitem [{\citenamefont {Yao}\ and\ \citenamefont {Wang}(2018)}]{yao2018}%
  \BibitemOpen
  \bibfield  {author} {\bibinfo {author} {\bibfnamefont {S.}~\bibnamefont {Yao}}\ and\ \bibinfo {author} {\bibfnamefont {Z.}~\bibnamefont {Wang}},\ }\bibfield  {title} {Edge States and Topological Invariants of Non-Hermitian Systems,\ }\href {https://link.aps.org/doi/10.1103/PhysRevLett.121.086803} {\bibfield  {journal} {\bibinfo  {journal} {Phys. Rev. Lett.}\ }\textbf {\bibinfo {volume} {121}},\ \bibinfo {pages} {086803} (\bibinfo {year} {2018})}\BibitemShut {NoStop}%
\bibitem [{\citenamefont {Kunst}\ \emph {et~al.}(2018)\citenamefont {Kunst}, \citenamefont {Edvardsson}, \citenamefont {Budich},\ and\ \citenamefont {Bergholtz}}]{kunst2018}%
  \BibitemOpen
  \bibfield  {author} {\bibinfo {author} {\bibfnamefont {F.~K.}\ \bibnamefont {Kunst}}, \bibinfo {author} {\bibfnamefont {E.}~\bibnamefont {Edvardsson}}, \bibinfo {author} {\bibfnamefont {J.~C.}\ \bibnamefont {Budich}}, \ and\ \bibinfo {author} {\bibfnamefont {E.~J.}\ \bibnamefont {Bergholtz}},\ }\bibfield  {title} {Biorthogonal Bulk-Boundary Correspondence in Non-Hermitian Systems,\ }\href {https://link.aps.org/doi/10.1103/PhysRevLett.121.026808} {\bibfield  {journal} {\bibinfo  {journal} {Phys. Rev. Lett.}\ }\textbf {\bibinfo {volume} {121}},\ \bibinfo {pages} {026808} (\bibinfo {year} {2018})}\BibitemShut {NoStop}%
\bibitem [{\citenamefont {Martinez~Alvarez}\ \emph {et~al.}(2018)\citenamefont {Martinez~Alvarez}, \citenamefont {Barrios~Vargas},\ and\ \citenamefont {Foa~Torres}}]{martinez2018}%
  \BibitemOpen
  \bibfield  {author} {\bibinfo {author} {\bibfnamefont {V.~M.}\ \bibnamefont {Martinez~Alvarez}}, \bibinfo {author} {\bibfnamefont {J.~E.}\ \bibnamefont {Barrios~Vargas}}, \ and\ \bibinfo {author} {\bibfnamefont {L.~E.~F.}\ \bibnamefont {Foa~Torres}},\ }\bibfield  {title} {Non-Hermitian robust edge states in one dimension: Anomalous localization and eigenspace condensation at exceptional points,\ }\href {https://link.aps.org/doi/10.1103/PhysRevB.97.121401} {\bibfield  {journal} {\bibinfo  {journal} {Phys. Rev. B}\ }\textbf {\bibinfo {volume} {97}},\ \bibinfo {pages} {121401} (\bibinfo {year} {2018})}\BibitemShut {NoStop}%
\bibitem [{\citenamefont {Lee}\ and\ \citenamefont {Thomale}(2019)}]{lee2019}%
  \BibitemOpen
  \bibfield  {author} {\bibinfo {author} {\bibfnamefont {C.~H.}\ \bibnamefont {Lee}}\ and\ \bibinfo {author} {\bibfnamefont {R.}~\bibnamefont {Thomale}},\ }\bibfield  {title} {Anatomy of skin modes and topology in non-Hermitian systems,\ }\href {https://link.aps.org/doi/10.1103/PhysRevB.99.201103} {\bibfield  {journal} {\bibinfo  {journal} {Phys. Rev. B}\ }\textbf {\bibinfo {volume} {99}},\ \bibinfo {pages} {201103} (\bibinfo {year} {2019})}\BibitemShut {NoStop}%
\bibitem [{\citenamefont {Borgnia}\ \emph {et~al.}(2020)\citenamefont {Borgnia}, \citenamefont {Kruchkov},\ and\ \citenamefont {Slager}}]{borgnia2020}%
  \BibitemOpen
  \bibfield  {author} {\bibinfo {author} {\bibfnamefont {D.~S.}\ \bibnamefont {Borgnia}}, \bibinfo {author} {\bibfnamefont {A.~J.}\ \bibnamefont {Kruchkov}}, \ and\ \bibinfo {author} {\bibfnamefont {R.-J.}\ \bibnamefont {Slager}},\ }\bibfield  {title} {Non-Hermitian Boundary Modes and Topology,\ }\href {https://link.aps.org/doi/10.1103/PhysRevLett.124.056802} {\bibfield  {journal} {\bibinfo  {journal} {Phys. Rev. Lett.}\ }\textbf {\bibinfo {volume} {124}},\ \bibinfo {pages} {056802} (\bibinfo {year} {2020})}\BibitemShut {NoStop}%
\bibitem [{\citenamefont {Zhang}\ \emph {et~al.}(2020)\citenamefont {Zhang}, \citenamefont {Yang},\ and\ \citenamefont {Fang}}]{zhang2020}%
  \BibitemOpen
  \bibfield  {author} {\bibinfo {author} {\bibfnamefont {K.}~\bibnamefont {Zhang}}, \bibinfo {author} {\bibfnamefont {Z.}~\bibnamefont {Yang}}, \ and\ \bibinfo {author} {\bibfnamefont {C.}~\bibnamefont {Fang}},\ }\bibfield  {title} {Correspondence between Winding Numbers and Skin Modes in Non-Hermitian Systems,\ }\href {https://link.aps.org/doi/10.1103/PhysRevLett.125.126402} {\bibfield  {journal} {\bibinfo  {journal} {Phys. Rev. Lett.}\ }\textbf {\bibinfo {volume} {125}},\ \bibinfo {pages} {126402} (\bibinfo {year} {2020})}\BibitemShut {NoStop}%
\bibitem [{\citenamefont {Okuma}\ \emph {et~al.}(2020)\citenamefont {Okuma}, \citenamefont {Kawabata}, \citenamefont {Shiozaki},\ and\ \citenamefont {Sato}}]{Okuma2020}%
  \BibitemOpen
  \bibfield  {author} {\bibinfo {author} {\bibfnamefont {N.}~\bibnamefont {Okuma}}, \bibinfo {author} {\bibfnamefont {K.}~\bibnamefont {Kawabata}}, \bibinfo {author} {\bibfnamefont {K.}~\bibnamefont {Shiozaki}}, \ and\ \bibinfo {author} {\bibfnamefont {M.}~\bibnamefont {Sato}},\ }\bibfield  {title} {Topological Origin of Non-Hermitian Skin Effects,\ }\href {https://link.aps.org/doi/10.1103/PhysRevLett.124.086801} {\bibfield  {journal} {\bibinfo  {journal} {Phys. Rev. Lett.}\ }\textbf {\bibinfo {volume} {124}},\ \bibinfo {pages} {086801} (\bibinfo {year} {2020})}\BibitemShut {NoStop}%
\bibitem [{\citenamefont {Okuma}\ and\ \citenamefont {Sato}(2023)}]{Okuma2023}%
  \BibitemOpen
  \bibfield  {author} {\bibinfo {author} {\bibfnamefont {N.}~\bibnamefont {Okuma}}\ and\ \bibinfo {author} {\bibfnamefont {M.}~\bibnamefont {Sato}},\ }\bibfield  {title} {Non-Hermitian Topological Phenomena: A Review,\ }\href {http://dx.doi.org/10.1146/annurev-conmatphys-040521-033133} {\bibfield  {journal} {\bibinfo  {journal} {Annual Review of Condensed Matter Physics}\ }\textbf {\bibinfo {volume} {14}},\ \bibinfo {pages} {83–107} (\bibinfo {year} {2023})}\BibitemShut {NoStop}%
\bibitem [{\citenamefont {Lin}\ \emph {et~al.}(2023)\citenamefont {Lin}, \citenamefont {Tai}, \citenamefont {Li},\ and\ \citenamefont {Lee}}]{Lin2023}%
  \BibitemOpen
  \bibfield  {author} {\bibinfo {author} {\bibfnamefont {R.}~\bibnamefont {Lin}}, \bibinfo {author} {\bibfnamefont {T.}~\bibnamefont {Tai}}, \bibinfo {author} {\bibfnamefont {L.}~\bibnamefont {Li}}, \ and\ \bibinfo {author} {\bibfnamefont {C.~H.}\ \bibnamefont {Lee}},\ }\bibfield  {title} {Topological non-Hermitian skin effect,\ }\href {http://dx.doi.org/10.1007/s11467-023-1309-z} {\bibfield  {journal} {\bibinfo  {journal} {Frontiers of Physics}\ }\textbf {\bibinfo {volume} {18}} (\bibinfo {year} {2023})}\BibitemShut {NoStop}%
\bibitem [{\citenamefont {Nelson}\ and\ \citenamefont {Vinokur}(1993)}]{Nelson1993}%
  \BibitemOpen
  \bibfield  {author} {\bibinfo {author} {\bibfnamefont {D.~R.}\ \bibnamefont {Nelson}}\ and\ \bibinfo {author} {\bibfnamefont {V.~M.}\ \bibnamefont {Vinokur}},\ }\bibfield  {title} {Boson localization and correlated pinning of superconducting vortex arrays,\ }\href {http://dx.doi.org/10.1103/PhysRevB.48.13060} {\bibfield  {journal} {\bibinfo  {journal} {Physical Review B}\ }\textbf {\bibinfo {volume} {48}},\ \bibinfo {pages} {13060–13097} (\bibinfo {year} {1993})}\BibitemShut {NoStop}%
\bibitem [{\citenamefont {Hatano}\ and\ \citenamefont {Nelson}(1996)}]{Hatano1996}%
  \BibitemOpen
  \bibfield  {author} {\bibinfo {author} {\bibfnamefont {N.}~\bibnamefont {Hatano}}\ and\ \bibinfo {author} {\bibfnamefont {D.~R.}\ \bibnamefont {Nelson}},\ }\bibfield  {title} {Localization Transitions in Non-Hermitian Quantum Mechanics,\ }\href {https://link.aps.org/doi/10.1103/PhysRevLett.77.570} {\bibfield  {journal} {\bibinfo  {journal} {Phys. Rev. Lett.}\ }\textbf {\bibinfo {volume} {77}},\ \bibinfo {pages} {570} (\bibinfo {year} {1996})}\BibitemShut {NoStop}%
\bibitem [{\citenamefont {Hatano}\ and\ \citenamefont {Nelson}(1997)}]{Hatano1997}%
  \BibitemOpen
  \bibfield  {author} {\bibinfo {author} {\bibfnamefont {N.}~\bibnamefont {Hatano}}\ and\ \bibinfo {author} {\bibfnamefont {D.~R.}\ \bibnamefont {Nelson}},\ }\bibfield  {title} {Vortex pinning and non-Hermitian quantum mechanics,\ }\href {https://link.aps.org/doi/10.1103/PhysRevB.56.8651} {\bibfield  {journal} {\bibinfo  {journal} {Phys. Rev. B}\ }\textbf {\bibinfo {volume} {56}},\ \bibinfo {pages} {8651} (\bibinfo {year} {1997})}\BibitemShut {NoStop}%
\bibitem [{\citenamefont {Hatano}\ and\ \citenamefont {Nelson}(1998)}]{Hatano1998}%
  \BibitemOpen
  \bibfield  {author} {\bibinfo {author} {\bibfnamefont {N.}~\bibnamefont {Hatano}}\ and\ \bibinfo {author} {\bibfnamefont {D.~R.}\ \bibnamefont {Nelson}},\ }\bibfield  {title} {Non-Hermitian delocalization and eigenfunctions,\ }\href {https://link.aps.org/doi/10.1103/PhysRevB.58.8384} {\bibfield  {journal} {\bibinfo  {journal} {Phys. Rev. B}\ }\textbf {\bibinfo {volume} {58}},\ \bibinfo {pages} {8384} (\bibinfo {year} {1998})}\BibitemShut {NoStop}%
\bibitem [{\citenamefont {Regensburger}\ \emph {et~al.}(2012)\citenamefont {Regensburger}, \citenamefont {Bersch}, \citenamefont {Miri}, \citenamefont {Onishchukov}, \citenamefont {Christodoulides},\ and\ \citenamefont {Peschel}}]{Regensburger2012}%
  \BibitemOpen
  \bibfield  {author} {\bibinfo {author} {\bibfnamefont {A.}~\bibnamefont {Regensburger}}, \bibinfo {author} {\bibfnamefont {C.}~\bibnamefont {Bersch}}, \bibinfo {author} {\bibfnamefont {M.-A.}\ \bibnamefont {Miri}}, \bibinfo {author} {\bibfnamefont {G.}~\bibnamefont {Onishchukov}}, \bibinfo {author} {\bibfnamefont {D.~N.}\ \bibnamefont {Christodoulides}}, \ and\ \bibinfo {author} {\bibfnamefont {U.}~\bibnamefont {Peschel}},\ }\bibfield  {title} {Parity–time synthetic photonic lattices,\ }\href {http://dx.doi.org/10.1038/nature11298} {\bibfield  {journal} {\bibinfo  {journal} {Nature}\ }\textbf {\bibinfo {volume} {488}},\ \bibinfo {pages} {167–171} (\bibinfo {year} {2012})}\BibitemShut {NoStop}%
\bibitem [{\citenamefont {Feng}\ \emph {et~al.}(2014)\citenamefont {Feng}, \citenamefont {Wong}, \citenamefont {Ma}, \citenamefont {Wang},\ and\ \citenamefont {Zhang}}]{Feng2014}%
  \BibitemOpen
  \bibfield  {author} {\bibinfo {author} {\bibfnamefont {L.}~\bibnamefont {Feng}}, \bibinfo {author} {\bibfnamefont {Z.~J.}\ \bibnamefont {Wong}}, \bibinfo {author} {\bibfnamefont {R.-M.}\ \bibnamefont {Ma}}, \bibinfo {author} {\bibfnamefont {Y.}~\bibnamefont {Wang}}, \ and\ \bibinfo {author} {\bibfnamefont {X.}~\bibnamefont {Zhang}},\ }\bibfield  {title} {Single-mode laser by parity-time symmetry breaking,\ }\href {http://dx.doi.org/10.1126/science.1258479} {\bibfield  {journal} {\bibinfo  {journal} {Science}\ }\textbf {\bibinfo {volume} {346}},\ \bibinfo {pages} {972–975} (\bibinfo {year} {2014})}\BibitemShut {NoStop}%
\bibitem [{\citenamefont {Wiersig}(2014)}]{Wiersig2014}%
  \BibitemOpen
  \bibfield  {author} {\bibinfo {author} {\bibfnamefont {J.}~\bibnamefont {Wiersig}},\ }\bibfield  {title} {Enhancing the Sensitivity of Frequency and Energy Splitting Detection by Using Exceptional Points: Application to Microcavity Sensors for Single-Particle Detection,\ }\href {http://dx.doi.org/10.1103/PhysRevLett.112.203901} {\bibfield  {journal} {\bibinfo  {journal} {Physical Review Letters}\ }\textbf {\bibinfo {volume} {112}} (\bibinfo {year} {2014})}\BibitemShut {NoStop}%
\bibitem [{\citenamefont {Longhi}\ \emph {et~al.}(2015)\citenamefont {Longhi}, \citenamefont {Gatti},\ and\ \citenamefont {Valle}}]{Longhi2015}%
  \BibitemOpen
  \bibfield  {author} {\bibinfo {author} {\bibfnamefont {S.}~\bibnamefont {Longhi}}, \bibinfo {author} {\bibfnamefont {D.}~\bibnamefont {Gatti}}, \ and\ \bibinfo {author} {\bibfnamefont {G.~D.}\ \bibnamefont {Valle}},\ }\bibfield  {title} {Robust light transport in non-Hermitian photonic lattices,\ }\href {http://dx.doi.org/10.1038/srep13376} {\bibfield  {journal} {\bibinfo  {journal} {Scientific Reports}\ }\textbf {\bibinfo {volume} {5}} (\bibinfo {year} {2015})}\BibitemShut {NoStop}%
\bibitem [{\citenamefont {Wiersig}(2016)}]{Wiersig2016}%
  \BibitemOpen
  \bibfield  {author} {\bibinfo {author} {\bibfnamefont {J.}~\bibnamefont {Wiersig}},\ }\bibfield  {title} {Sensors operating at exceptional points: General theory,\ }\href {http://dx.doi.org/10.1103/PhysRevA.93.033809} {\bibfield  {journal} {\bibinfo  {journal} {Physical Review A}\ }\textbf {\bibinfo {volume} {93}} (\bibinfo {year} {2016})}\BibitemShut {NoStop}%
\bibitem [{\citenamefont {Liu}\ \emph {et~al.}(2016)\citenamefont {Liu}, \citenamefont {Zhang}, \citenamefont {\"Ozdemir}, \citenamefont {Peng}, \citenamefont {Jing}, \citenamefont {L\"u}, \citenamefont {Li}, \citenamefont {Yang}, \citenamefont {Nori},\ and\ \citenamefont {Liu}}]{Liu2016}%
  \BibitemOpen
  \bibfield  {author} {\bibinfo {author} {\bibfnamefont {Z.-P.}\ \bibnamefont {Liu}}, \bibinfo {author} {\bibfnamefont {J.}~\bibnamefont {Zhang}}, \bibinfo {author} {\bibfnamefont {i.~m. c.~K.}\ \bibnamefont {\"Ozdemir}}, \bibinfo {author} {\bibfnamefont {B.}~\bibnamefont {Peng}}, \bibinfo {author} {\bibfnamefont {H.}~\bibnamefont {Jing}}, \bibinfo {author} {\bibfnamefont {X.-Y.}\ \bibnamefont {L\"u}}, \bibinfo {author} {\bibfnamefont {C.-W.}\ \bibnamefont {Li}}, \bibinfo {author} {\bibfnamefont {L.}~\bibnamefont {Yang}}, \bibinfo {author} {\bibfnamefont {F.}~\bibnamefont {Nori}}, \ and\ \bibinfo {author} {\bibfnamefont {Y.-x.}\ \bibnamefont {Liu}},\ }\bibfield  {title} {Metrology with PT-Symmetric Cavities: Enhanced Sensitivity near the PT-Phase Transition,\ }\href {\doibase/10.1103/PhysRevLett.117.110802} {\bibfield  {journal} {\bibinfo  {journal} {Phys. Rev. Lett.}\ }\textbf {\bibinfo {volume} {117}},\ \bibinfo {pages} {110802} (\bibinfo {year} {2016})}\BibitemShut {NoStop}%
\bibitem [{\citenamefont {Chen}\ \emph {et~al.}(2017)\citenamefont {Chen}, \citenamefont {Kaya~\"Ozdemir}, \citenamefont {Zhao}, \citenamefont {Wiersig},\ and\ \citenamefont {Yang}}]{Chen2017}%
  \BibitemOpen
  \bibfield  {author} {\bibinfo {author} {\bibfnamefont {W.}~\bibnamefont {Chen}}, \bibinfo {author} {\bibfnamefont {S.}~\bibnamefont {Kaya~\"Ozdemir}}, \bibinfo {author} {\bibfnamefont {G.}~\bibnamefont {Zhao}}, \bibinfo {author} {\bibfnamefont {J.}~\bibnamefont {Wiersig}}, \ and\ \bibinfo {author} {\bibfnamefont {L.}~\bibnamefont {Yang}},\ }\bibfield  {title} {Exceptional points enhance sensing in an optical microcavity,\ }\href {\doibase/10.1038/nature23281} {\bibfield  {journal} {\bibinfo  {journal} {Nature}\ }\textbf {\bibinfo {volume} {548}},\ \bibinfo {pages} {192–196} (\bibinfo {year} {2017})}\BibitemShut {NoStop}%
\bibitem [{\citenamefont {Weidemann}\ \emph {et~al.}(2020)\citenamefont {Weidemann}, \citenamefont {Kremer}, \citenamefont {Helbig}, \citenamefont {Hofmann}, \citenamefont {Stegmaier}, \citenamefont {Greiter}, \citenamefont {Thomale},\ and\ \citenamefont {Szameit}}]{Weidemann2020}%
  \BibitemOpen
  \bibfield  {author} {\bibinfo {author} {\bibfnamefont {S.}~\bibnamefont {Weidemann}}, \bibinfo {author} {\bibfnamefont {M.}~\bibnamefont {Kremer}}, \bibinfo {author} {\bibfnamefont {T.}~\bibnamefont {Helbig}}, \bibinfo {author} {\bibfnamefont {T.}~\bibnamefont {Hofmann}}, \bibinfo {author} {\bibfnamefont {A.}~\bibnamefont {Stegmaier}}, \bibinfo {author} {\bibfnamefont {M.}~\bibnamefont {Greiter}}, \bibinfo {author} {\bibfnamefont {R.}~\bibnamefont {Thomale}}, \ and\ \bibinfo {author} {\bibfnamefont {A.}~\bibnamefont {Szameit}},\ }\bibfield  {title} {Topological funneling of light,\ }\href {http://dx.doi.org/10.1126/science.aaz8727} {\bibfield  {journal} {\bibinfo  {journal} {Science}\ }\textbf {\bibinfo {volume} {368}},\ \bibinfo {pages} {311–314} (\bibinfo {year} {2020})}\BibitemShut {NoStop}%
\bibitem [{\citenamefont {Zhang}\ \emph {et~al.}(2021)\citenamefont {Zhang}, \citenamefont {Lv}, \citenamefont {Yan},\ and\ \citenamefont {Zhou}}]{Zhang2021}%
  \BibitemOpen
  \bibfield  {author} {\bibinfo {author} {\bibfnamefont {R.}~\bibnamefont {Zhang}}, \bibinfo {author} {\bibfnamefont {C.}~\bibnamefont {Lv}}, \bibinfo {author} {\bibfnamefont {Y.}~\bibnamefont {Yan}}, \ and\ \bibinfo {author} {\bibfnamefont {Q.}~\bibnamefont {Zhou}},\ }\bibfield  {title} {Efimov-like states and quantum funneling effects on synthetic hyperbolic surfaces,\ }\href {http://dx.doi.org/10.1016/j.scib.2021.06.017} {\bibfield  {journal} {\bibinfo  {journal} {Science Bulletin}\ }\textbf {\bibinfo {volume} {66}},\ \bibinfo {pages} {1967–1972} (\bibinfo {year} {2021})}\BibitemShut {NoStop}%
\bibitem [{\citenamefont {Hughes}\ and\ \citenamefont {Sahimi}(1982)}]{Hughes1982}%
  \BibitemOpen
  \bibfield  {author} {\bibinfo {author} {\bibfnamefont {B.~D.}\ \bibnamefont {Hughes}}\ and\ \bibinfo {author} {\bibfnamefont {M.}~\bibnamefont {Sahimi}},\ }\bibfield  {title} {Random walks on the Bethe lattice,\ }\href {\doibase/10.1007/bf01011791} {\bibfield  {journal} {\bibinfo  {journal} {Journal of Statistical Physics}\ }\textbf {\bibinfo {volume} {29}},\ \bibinfo {pages} {781–794} (\bibinfo {year} {1982})}\BibitemShut {NoStop}%
\bibitem [{\citenamefont {Cassi}(1990)}]{Cassi1990}%
  \BibitemOpen
  \bibfield  {author} {\bibinfo {author} {\bibfnamefont {D.}~\bibnamefont {Cassi}},\ }\bibfield  {title} {Dynamical Phase Transition of Biased Random Walks on Bethe Lattices,\ }\href {\doibase/10.1209/0295-5075/13/7/002} {\bibfield  {journal} {\bibinfo  {journal} {Europhysics Letters (EPL)}\ }\textbf {\bibinfo {volume} {13}},\ \bibinfo {pages} {583–586} (\bibinfo {year} {1990})}\BibitemShut {NoStop}%
\bibitem [{\citenamefont {Monthus}\ and\ \citenamefont {Texier}(1996)}]{Monthus1996}%
  \BibitemOpen
  \bibfield  {author} {\bibinfo {author} {\bibfnamefont {C.}~\bibnamefont {Monthus}}\ and\ \bibinfo {author} {\bibfnamefont {C.}~\bibnamefont {Texier}},\ }\bibfield  {title} {Random walk on the Bethe lattice and hyperbolic Brownian motion,\ }\href {\doibase/10.1088/0305-4470/29/10/019} {\bibfield  {journal} {\bibinfo  {journal} {Journal of Physics A: Mathematical and General}\ }\textbf {\bibinfo {volume} {29}},\ \bibinfo {pages} {2399–2409} (\bibinfo {year} {1996})}\BibitemShut {NoStop}%
\bibitem [{\citenamefont {Lv}\ \emph {et~al.}(2022)\citenamefont {Lv}, \citenamefont {Zhang}, \citenamefont {Zhai},\ and\ \citenamefont {Zhou}}]{Lv2022}%
  \BibitemOpen
  \bibfield  {author} {\bibinfo {author} {\bibfnamefont {C.}~\bibnamefont {Lv}}, \bibinfo {author} {\bibfnamefont {R.}~\bibnamefont {Zhang}}, \bibinfo {author} {\bibfnamefont {Z.}~\bibnamefont {Zhai}}, \ and\ \bibinfo {author} {\bibfnamefont {Q.}~\bibnamefont {Zhou}},\ }\bibfield  {title} {Curving the space by non-Hermiticity,\ }\href {http://dx.doi.org/10.1038/s41467-022-29774-8} {\bibfield  {journal} {\bibinfo  {journal} {Nature Communications}\ }\textbf {\bibinfo {volume} {13}} (\bibinfo {year} {2022})}\BibitemShut {NoStop}%
\bibitem [{\citenamefont {da~Costa}(1981)}]{daCosta1982}%
  \BibitemOpen
  \bibfield  {author} {\bibinfo {author} {\bibfnamefont {R.~C.~T.}\ \bibnamefont {da~Costa}},\ }\bibfield  {title} {Quantum mechanics of a constrained particle,\ }\href {https://link.aps.org/doi/10.1103/PhysRevA.23.1982} {\bibfield  {journal} {\bibinfo  {journal} {Phys. Rev. A}\ }\textbf {\bibinfo {volume} {23}},\ \bibinfo {pages} {1982} (\bibinfo {year} {1981})}\BibitemShut {NoStop}%
\bibitem [{\citenamefont {Li}\ \emph {et~al.}(2020)\citenamefont {Li}, \citenamefont {Lee}, \citenamefont {Mu},\ and\ \citenamefont {Gong}}]{Li2020}%
  \BibitemOpen
  \bibfield  {author} {\bibinfo {author} {\bibfnamefont {L.}~\bibnamefont {Li}}, \bibinfo {author} {\bibfnamefont {C.~H.}\ \bibnamefont {Lee}}, \bibinfo {author} {\bibfnamefont {S.}~\bibnamefont {Mu}}, \ and\ \bibinfo {author} {\bibfnamefont {J.}~\bibnamefont {Gong}},\ }\bibfield  {title} {Critical non-Hermitian skin effect,\ }\href {http://dx.doi.org/10.1038/s41467-020-18917-4} {\bibfield  {journal} {\bibinfo  {journal} {Nature Communications}\ }\textbf {\bibinfo {volume} {11}} (\bibinfo {year} {2020})}\BibitemShut {NoStop}%
\bibitem [{\citenamefont {Yang}\ \emph {et~al.}(2021)\citenamefont {Yang}, \citenamefont {Morampudi},\ and\ \citenamefont {Bergholtz}}]{Yang2021}%
  \BibitemOpen
  \bibfield  {author} {\bibinfo {author} {\bibfnamefont {K.}~\bibnamefont {Yang}}, \bibinfo {author} {\bibfnamefont {S.~C.}\ \bibnamefont {Morampudi}}, \ and\ \bibinfo {author} {\bibfnamefont {E.~J.}\ \bibnamefont {Bergholtz}},\ }\bibfield  {title} {Exceptional Spin Liquids from Couplings to the Environment,\ }\href {http://dx.doi.org/10.1103/PhysRevLett.126.077201} {\bibfield  {journal} {\bibinfo  {journal} {Physical Review Letters}\ }\textbf {\bibinfo {volume} {126}} (\bibinfo {year} {2021})}\BibitemShut {NoStop}%
\bibitem [{\citenamefont {Shen}\ and\ \citenamefont {Lee}(2022)}]{Shen2022}%
  \BibitemOpen
  \bibfield  {author} {\bibinfo {author} {\bibfnamefont {R.}~\bibnamefont {Shen}}\ and\ \bibinfo {author} {\bibfnamefont {C.~H.}\ \bibnamefont {Lee}},\ }\bibfield  {title} {Non-Hermitian skin clusters from strong interactions,\ }\href {http://dx.doi.org/10.1038/s42005-022-01015-w} {\bibfield  {journal} {\bibinfo  {journal} {Communications Physics}\ }\textbf {\bibinfo {volume} {5}} (\bibinfo {year} {2022})}\BibitemShut {NoStop}%
\bibitem [{\citenamefont {Lee}(2021)}]{Lee2021}%
  \BibitemOpen
  \bibfield  {author} {\bibinfo {author} {\bibfnamefont {C.~H.}\ \bibnamefont {Lee}},\ }\bibfield  {title} {Many-body topological and skin states without open boundaries,\ }\href {http://dx.doi.org/10.1103/PhysRevB.104.195102} {\bibfield  {journal} {\bibinfo  {journal} {Physical Review B}\ }\textbf {\bibinfo {volume} {104}} (\bibinfo {year} {2021})}\BibitemShut {NoStop}%
\bibitem [{\citenamefont {Kawabata}\ \emph {et~al.}(2022)\citenamefont {Kawabata}, \citenamefont {Shiozaki},\ and\ \citenamefont {Ryu}}]{Kawabata2022}%
  \BibitemOpen
  \bibfield  {author} {\bibinfo {author} {\bibfnamefont {K.}~\bibnamefont {Kawabata}}, \bibinfo {author} {\bibfnamefont {K.}~\bibnamefont {Shiozaki}}, \ and\ \bibinfo {author} {\bibfnamefont {S.}~\bibnamefont {Ryu}},\ }\bibfield  {title} {Many-body topology of non-Hermitian systems,\ }\href {http://dx.doi.org/10.1103/PhysRevB.105.165137} {\bibfield  {journal} {\bibinfo  {journal} {Physical Review B}\ }\textbf {\bibinfo {volume} {105}} (\bibinfo {year} {2022})}\BibitemShut {NoStop}%
\bibitem [{\citenamefont {Alsallom}\ \emph {et~al.}(2022)\citenamefont {Alsallom}, \citenamefont {Herviou}, \citenamefont {Yazyev},\ and\ \citenamefont {Brzezińska}}]{Alsallom2022}%
  \BibitemOpen
  \bibfield  {author} {\bibinfo {author} {\bibfnamefont {F.}~\bibnamefont {Alsallom}}, \bibinfo {author} {\bibfnamefont {L.}~\bibnamefont {Herviou}}, \bibinfo {author} {\bibfnamefont {O.~V.}\ \bibnamefont {Yazyev}}, \ and\ \bibinfo {author} {\bibfnamefont {M.}~\bibnamefont {Brzezińska}},\ }\bibfield  {title} {Fate of the non-Hermitian skin effect in many-body fermionic systems,\ }\href {http://dx.doi.org/10.1103/PhysRevResearch.4.033122} {\bibfield  {journal} {\bibinfo  {journal} {Physical Review Research}\ }\textbf {\bibinfo {volume} {4}} (\bibinfo {year} {2022})}\BibitemShut {NoStop}%
\bibitem [{\citenamefont {Gliozzi}\ \emph {et~al.}(2024)\citenamefont {Gliozzi}, \citenamefont {De~Tomasi},\ and\ \citenamefont {Hughes}}]{Gliozzi2024}%
  \BibitemOpen
  \bibfield  {author} {\bibinfo {author} {\bibfnamefont {J.}~\bibnamefont {Gliozzi}}, \bibinfo {author} {\bibfnamefont {G.}~\bibnamefont {De~Tomasi}}, \ and\ \bibinfo {author} {\bibfnamefont {T.~L.}\ \bibnamefont {Hughes}},\ }\bibfield  {title} {Many-Body Non-Hermitian Skin Effect for Multipoles,\ }\href {http://dx.doi.org/10.1103/PhysRevLett.133.136503} {\bibfield  {journal} {\bibinfo  {journal} {Physical Review Letters}\ }\textbf {\bibinfo {volume} {133}} (\bibinfo {year} {2024})}\BibitemShut {NoStop}%
\bibitem [{\citenamefont {Pretko}(2017)}]{Pretko2017}%
  \BibitemOpen
  \bibfield  {author} {\bibinfo {author} {\bibfnamefont {M.}~\bibnamefont {Pretko}},\ }\bibfield  {title} {Subdimensional particle structure of higher rank $U(1)$ spin liquids,\ }\href {http://dx.doi.org/10.1103/PhysRevB.95.115139} {\bibfield  {journal} {\bibinfo  {journal} {Physical Review B}\ }\textbf {\bibinfo {volume} {95}} (\bibinfo {year} {2017})}\BibitemShut {NoStop}%
\bibitem [{\citenamefont {Ma}\ \emph {et~al.}(2018)\citenamefont {Ma}, \citenamefont {Hermele},\ and\ \citenamefont {Chen}}]{Ma2018}%
  \BibitemOpen
  \bibfield  {author} {\bibinfo {author} {\bibfnamefont {H.}~\bibnamefont {Ma}}, \bibinfo {author} {\bibfnamefont {M.}~\bibnamefont {Hermele}}, \ and\ \bibinfo {author} {\bibfnamefont {X.}~\bibnamefont {Chen}},\ }\bibfield  {title} {Fracton topological order from the Higgs and partial-confinement mechanisms of rank-two gauge theory,\ }\href {http://dx.doi.org/10.1103/PhysRevB.98.035111} {\bibfield  {journal} {\bibinfo  {journal} {Physical Review B}\ }\textbf {\bibinfo {volume} {98}} (\bibinfo {year} {2018})}\BibitemShut {NoStop}%
\bibitem [{\citenamefont {Bulmash}\ and\ \citenamefont {Barkeshli}(2018)}]{Bulmash2018}%
  \BibitemOpen
  \bibfield  {author} {\bibinfo {author} {\bibfnamefont {D.}~\bibnamefont {Bulmash}}\ and\ \bibinfo {author} {\bibfnamefont {M.}~\bibnamefont {Barkeshli}},\ }\bibfield  {title} {Higgs mechanism in higher-rank symmetric U(1) gauge theories,\ }\href {http://dx.doi.org/10.1103/PhysRevB.97.235112} {\bibfield  {journal} {\bibinfo  {journal} {Physical Review B}\ }\textbf {\bibinfo {volume} {97}} (\bibinfo {year} {2018})}\BibitemShut {NoStop}%
\bibitem [{\citenamefont {Zhang}\ \emph {et~al.}(2025)\citenamefont {Zhang}, \citenamefont {Lv},\ and\ \citenamefont {Zhou}}]{Zhang2025}%
  \BibitemOpen
  \bibfield  {author} {\bibinfo {author} {\bibfnamefont {S.}~\bibnamefont {Zhang}}, \bibinfo {author} {\bibfnamefont {C.}~\bibnamefont {Lv}}, \ and\ \bibinfo {author} {\bibfnamefont {Q.}~\bibnamefont {Zhou}},\ }\bibfield  {title} {Synthetic tensor gauge fields,\ }\href {http://dx.doi.org/10.1103/PhysRevResearch.7.013013} {\bibfield  {journal} {\bibinfo  {journal} {Physical Review Research}\ }\textbf {\bibinfo {volume} {7}} (\bibinfo {year} {2025})}\BibitemShut {NoStop}%
\bibitem [{\citenamefont {Lin}\ \emph {et~al.}(2011)\citenamefont {Lin}, \citenamefont {Jiménez-García},\ and\ \citenamefont {Spielman}}]{Lin2011}%
  \BibitemOpen
  \bibfield  {author} {\bibinfo {author} {\bibfnamefont {Y.-J.}\ \bibnamefont {Lin}}, \bibinfo {author} {\bibfnamefont {K.}~\bibnamefont {Jiménez-García}}, \ and\ \bibinfo {author} {\bibfnamefont {I.~B.}\ \bibnamefont {Spielman}},\ }\bibfield  {title} {Spin–orbit-coupled Bose–Einstein condensates,\ }\href {http://dx.doi.org/10.1038/nature09887} {\bibfield  {journal} {\bibinfo  {journal} {Nature}\ }\textbf {\bibinfo {volume} {471}},\ \bibinfo {pages} {83–86} (\bibinfo {year} {2011})}\BibitemShut {NoStop}%
\bibitem [{\citenamefont {Galitski}\ and\ \citenamefont {Spielman}(2013)}]{Galitski2013}%
  \BibitemOpen
  \bibfield  {author} {\bibinfo {author} {\bibfnamefont {V.}~\bibnamefont {Galitski}}\ and\ \bibinfo {author} {\bibfnamefont {I.~B.}\ \bibnamefont {Spielman}},\ }\bibfield  {title} {Spin–orbit coupling in quantum gases,\ }\href {http://dx.doi.org/10.1038/nature11841} {\bibfield  {journal} {\bibinfo  {journal} {Nature}\ }\textbf {\bibinfo {volume} {494}},\ \bibinfo {pages} {49–54} (\bibinfo {year} {2013})}\BibitemShut {NoStop}%
\bibitem [{\citenamefont {Ren}\ \emph {et~al.}(2022)\citenamefont {Ren}, \citenamefont {Liu}, \citenamefont {Zhao}, \citenamefont {He}, \citenamefont {Pak}, \citenamefont {Li},\ and\ \citenamefont {Jo}}]{Ren2022}%
  \BibitemOpen
  \bibfield  {author} {\bibinfo {author} {\bibfnamefont {Z.}~\bibnamefont {Ren}}, \bibinfo {author} {\bibfnamefont {D.}~\bibnamefont {Liu}}, \bibinfo {author} {\bibfnamefont {E.}~\bibnamefont {Zhao}}, \bibinfo {author} {\bibfnamefont {C.}~\bibnamefont {He}}, \bibinfo {author} {\bibfnamefont {K.~K.}\ \bibnamefont {Pak}}, \bibinfo {author} {\bibfnamefont {J.}~\bibnamefont {Li}}, \ and\ \bibinfo {author} {\bibfnamefont {G.-B.}\ \bibnamefont {Jo}},\ }\bibfield  {title} {Chiral control of quantum states in non-Hermitian spin–orbit-coupled fermions,\ }\href {http://dx.doi.org/10.1038/s41567-021-01491-x} {\bibfield  {journal} {\bibinfo  {journal} {Nature Physics}\ }\textbf {\bibinfo {volume} {18}},\ \bibinfo {pages} {385–389} (\bibinfo {year} {2022})}\BibitemShut {NoStop}%
\bibitem [{\citenamefont {Tao}\ \emph {et~al.}(2025)\citenamefont {Tao}, \citenamefont {Mercado-Gutierrez}, \citenamefont {Zhao},\ and\ \citenamefont {Spielman}}]{Tao2025}%
  \BibitemOpen
  \bibfield  {author} {\bibinfo {author} {\bibfnamefont {J.}~\bibnamefont {Tao}}, \bibinfo {author} {\bibfnamefont {E.}~\bibnamefont {Mercado-Gutierrez}}, \bibinfo {author} {\bibfnamefont {M.}~\bibnamefont {Zhao}}, \ and\ \bibinfo {author} {\bibfnamefont {I.}~\bibnamefont {Spielman}},\ }\href {https://arxiv.org/abs/2504.08614} {Imaginary gauge potentials in a non-Hermitian spin-orbit coupled quantum gas} (\bibinfo {year} {2025})\BibitemShut {NoStop}%
\bibitem [{\citenamefont {Zhou}\ \emph {et~al.}(2022)\citenamefont {Zhou}, \citenamefont {Li}, \citenamefont {Yi},\ and\ \citenamefont {Cui}}]{Zhou2022}%
  \BibitemOpen
  \bibfield  {author} {\bibinfo {author} {\bibfnamefont {L.}~\bibnamefont {Zhou}}, \bibinfo {author} {\bibfnamefont {H.}~\bibnamefont {Li}}, \bibinfo {author} {\bibfnamefont {W.}~\bibnamefont {Yi}}, \ and\ \bibinfo {author} {\bibfnamefont {X.}~\bibnamefont {Cui}},\ }\bibfield  {title} {Engineering non-Hermitian skin effect with band topology in ultracold gases,\ }\href {http://dx.doi.org/10.1038/s42005-022-01021-y} {\bibfield  {journal} {\bibinfo  {journal} {Communications Physics}\ }\textbf {\bibinfo {volume} {5}} (\bibinfo {year} {2022})}\BibitemShut {NoStop}%
\bibitem [{\citenamefont {Griesmaier}\ \emph {et~al.}(2005)\citenamefont {Griesmaier}, \citenamefont {Werner}, \citenamefont {Hensler}, \citenamefont {Stuhler},\ and\ \citenamefont {Pfau}}]{Griesmaier2005}%
  \BibitemOpen
  \bibfield  {author} {\bibinfo {author} {\bibfnamefont {A.}~\bibnamefont {Griesmaier}}, \bibinfo {author} {\bibfnamefont {J.}~\bibnamefont {Werner}}, \bibinfo {author} {\bibfnamefont {S.}~\bibnamefont {Hensler}}, \bibinfo {author} {\bibfnamefont {J.}~\bibnamefont {Stuhler}}, \ and\ \bibinfo {author} {\bibfnamefont {T.}~\bibnamefont {Pfau}},\ }\bibfield  {title} {Bose-Einstein Condensation of Chromium,\ }\href {\doibase/10.1103/PhysRevLett.94.160401} {\bibfield  {journal} {\bibinfo  {journal} {Phys. Rev. Lett.}\ }\textbf {\bibinfo {volume} {94}},\ \bibinfo {pages} {160401} (\bibinfo {year} {2005})}\BibitemShut {NoStop}%
\bibitem [{\citenamefont {Stuhler}\ \emph {et~al.}(2005)\citenamefont {Stuhler}, \citenamefont {Griesmaier}, \citenamefont {Koch}, \citenamefont {Fattori}, \citenamefont {Pfau}, \citenamefont {Giovanazzi}, \citenamefont {Pedri},\ and\ \citenamefont {Santos}}]{Stuhler2005}%
  \BibitemOpen
  \bibfield  {author} {\bibinfo {author} {\bibfnamefont {J.}~\bibnamefont {Stuhler}}, \bibinfo {author} {\bibfnamefont {A.}~\bibnamefont {Griesmaier}}, \bibinfo {author} {\bibfnamefont {T.}~\bibnamefont {Koch}}, \bibinfo {author} {\bibfnamefont {M.}~\bibnamefont {Fattori}}, \bibinfo {author} {\bibfnamefont {T.}~\bibnamefont {Pfau}}, \bibinfo {author} {\bibfnamefont {S.}~\bibnamefont {Giovanazzi}}, \bibinfo {author} {\bibfnamefont {P.}~\bibnamefont {Pedri}}, \ and\ \bibinfo {author} {\bibfnamefont {L.}~\bibnamefont {Santos}},\ }\bibfield  {title} {Observation of Dipole-Dipole Interaction in a Degenerate Quantum Gas,\ }\href {https://link.aps.org/doi/10.1103/PhysRevLett.95.150406} {\bibfield  {journal} {\bibinfo  {journal} {Phys. Rev. Lett.}\ }\textbf {\bibinfo {volume} {95}},\ \bibinfo {pages} {150406} (\bibinfo {year} {2005})}\BibitemShut {NoStop}%
\bibitem [{\citenamefont {Gallagher}\ and\ \citenamefont {Pillet}(2008)}]{Gallagher2008}%
  \BibitemOpen
  \bibfield  {author} {\bibinfo {author} {\bibfnamefont {T.~F.}\ \bibnamefont {Gallagher}}\ and\ \bibinfo {author} {\bibfnamefont {P.}~\bibnamefont {Pillet}},\ }Dipole–Dipole Interactions of Rydberg Atoms,\ in\ \href {http://dx.doi.org/10.1016/S1049-250X(08)00013-X} {\emph {\bibinfo {booktitle} {Advances in Atomic, Molecular, and Optical Physics}}}\ (\bibinfo  {publisher} {Elsevier},\ \bibinfo {year} {2008})\ p.\ \bibinfo {pages} {161–218}\BibitemShut {NoStop}%
\bibitem [{\citenamefont {Browaeys}\ \emph {et~al.}(2016)\citenamefont {Browaeys}, \citenamefont {Barredo},\ and\ \citenamefont {Lahaye}}]{Browaeys2016}%
  \BibitemOpen
  \bibfield  {author} {\bibinfo {author} {\bibfnamefont {A.}~\bibnamefont {Browaeys}}, \bibinfo {author} {\bibfnamefont {D.}~\bibnamefont {Barredo}}, \ and\ \bibinfo {author} {\bibfnamefont {T.}~\bibnamefont {Lahaye}},\ }\bibfield  {title} {Experimental investigations of dipole–dipole interactions between a few Rydberg atoms,\ }\href {http://dx.doi.org/10.1088/0953-4075/49/15/152001} {\bibfield  {journal} {\bibinfo  {journal} {Journal of Physics B: Atomic, Molecular and Optical Physics}\ }\textbf {\bibinfo {volume} {49}},\ \bibinfo {pages} {152001} (\bibinfo {year} {2016})}\BibitemShut {NoStop}%
\bibitem [{\citenamefont {Yoshida}(2013)}]{Yoshida2013}%
  \BibitemOpen
  \bibfield  {author} {\bibinfo {author} {\bibfnamefont {B.}~\bibnamefont {Yoshida}},\ }\bibfield  {title} {Exotic topological order in fractal spin liquids,\ }\href {\doibase/10.1103/PhysRevB.88.125122} {\bibfield  {journal} {\bibinfo  {journal} {Phys. Rev. B}\ }\textbf {\bibinfo {volume} {88}},\ \bibinfo {pages} {125122} (\bibinfo {year} {2013})}\BibitemShut {NoStop}%
\bibitem [{\citenamefont {Vijay}\ \emph {et~al.}(2015)\citenamefont {Vijay}, \citenamefont {Haah},\ and\ \citenamefont {Fu}}]{Vijay2015}%
  \BibitemOpen
  \bibfield  {author} {\bibinfo {author} {\bibfnamefont {S.}~\bibnamefont {Vijay}}, \bibinfo {author} {\bibfnamefont {J.}~\bibnamefont {Haah}}, \ and\ \bibinfo {author} {\bibfnamefont {L.}~\bibnamefont {Fu}},\ }\bibfield  {title} {A new kind of topological quantum order: A dimensional hierarchy of quasiparticles built from stationary excitations,\ }\href {\doibase/10.1103/PhysRevB.92.235136} {\bibfield  {journal} {\bibinfo  {journal} {Phys. Rev. B}\ }\textbf {\bibinfo {volume} {92}},\ \bibinfo {pages} {235136} (\bibinfo {year} {2015})}\BibitemShut {NoStop}%
\bibitem [{\citenamefont {Vijay}\ \emph {et~al.}(2016)\citenamefont {Vijay}, \citenamefont {Haah},\ and\ \citenamefont {Fu}}]{Vijay2016}%
  \BibitemOpen
  \bibfield  {author} {\bibinfo {author} {\bibfnamefont {S.}~\bibnamefont {Vijay}}, \bibinfo {author} {\bibfnamefont {J.}~\bibnamefont {Haah}}, \ and\ \bibinfo {author} {\bibfnamefont {L.}~\bibnamefont {Fu}},\ }\bibfield  {title} {Fracton topological order, generalized lattice gauge theory, and duality,\ }\href {\doibase/10.1103/PhysRevB.94.235157} {\bibfield  {journal} {\bibinfo  {journal} {Phys. Rev. B}\ }\textbf {\bibinfo {volume} {94}},\ \bibinfo {pages} {235157} (\bibinfo {year} {2016})}\BibitemShut {NoStop}%
\bibitem [{\citenamefont {Prem}\ \emph {et~al.}(2018)\citenamefont {Prem}, \citenamefont {Pretko},\ and\ \citenamefont {Nandkishore}}]{Prem2018}%
  \BibitemOpen
  \bibfield  {author} {\bibinfo {author} {\bibfnamefont {A.}~\bibnamefont {Prem}}, \bibinfo {author} {\bibfnamefont {M.}~\bibnamefont {Pretko}}, \ and\ \bibinfo {author} {\bibfnamefont {R.~M.}\ \bibnamefont {Nandkishore}},\ }\bibfield  {title} {Emergent phases of fractonic matter,\ }\href {\doibase/10.1103/PhysRevB.97.085116} {\bibfield  {journal} {\bibinfo  {journal} {Phys. Rev. B}\ }\textbf {\bibinfo {volume} {97}},\ \bibinfo {pages} {085116} (\bibinfo {year} {2018})}\BibitemShut {NoStop}%
\bibitem [{\citenamefont {Nandkishore}\ and\ \citenamefont {Hermele}(2019)}]{Nandkishore2019}%
  \BibitemOpen
  \bibfield  {author} {\bibinfo {author} {\bibfnamefont {R.~M.}\ \bibnamefont {Nandkishore}}\ and\ \bibinfo {author} {\bibfnamefont {M.}~\bibnamefont {Hermele}},\ }\bibfield  {title} {Fractons,\ }\href {http://dx.doi.org/10.1146/annurev-conmatphys-031218-013604} {\bibfield  {journal} {\bibinfo  {journal} {Annual Review of Condensed Matter Physics}\ }\textbf {\bibinfo {volume} {10}},\ \bibinfo {pages} {295–313} (\bibinfo {year} {2019})}\BibitemShut {NoStop}%
\bibitem [{\citenamefont {Pretko}\ \emph {et~al.}(2020)\citenamefont {Pretko}, \citenamefont {Chen},\ and\ \citenamefont {You}}]{Pretko2020}%
  \BibitemOpen
  \bibfield  {author} {\bibinfo {author} {\bibfnamefont {M.}~\bibnamefont {Pretko}}, \bibinfo {author} {\bibfnamefont {X.}~\bibnamefont {Chen}}, \ and\ \bibinfo {author} {\bibfnamefont {Y.}~\bibnamefont {You}},\ }\bibfield  {title} {Fracton phases of matter,\ }\href {\doibase/10.1142/s0217751x20300033} {\bibfield  {journal} {\bibinfo  {journal} {International Journal of Modern Physics A}\ }\textbf {\bibinfo {volume} {35}},\ \bibinfo {pages} {2030003} (\bibinfo {year} {2020})}\BibitemShut {NoStop}%
\bibitem [{\citenamefont {Yuan}\ \emph {et~al.}(2020)\citenamefont {Yuan}, \citenamefont {Chen},\ and\ \citenamefont {Ye}}]{Yuan2020}%
  \BibitemOpen
  \bibfield  {author} {\bibinfo {author} {\bibfnamefont {J.-K.}\ \bibnamefont {Yuan}}, \bibinfo {author} {\bibfnamefont {S.~A.}\ \bibnamefont {Chen}}, \ and\ \bibinfo {author} {\bibfnamefont {P.}~\bibnamefont {Ye}},\ }\bibfield  {title} {Fractonic superfluids,\ }\href {\doibase/10.1103/PhysRevResearch.2.023267} {\bibfield  {journal} {\bibinfo  {journal} {Phys. Rev. Res.}\ }\textbf {\bibinfo {volume} {2}},\ \bibinfo {pages} {023267} (\bibinfo {year} {2020})}\BibitemShut {NoStop}%
\bibitem [{\citenamefont {Lake}\ \emph {et~al.}(2022)\citenamefont {Lake}, \citenamefont {Hermele},\ and\ \citenamefont {Senthil}}]{Lake2022}%
  \BibitemOpen
  \bibfield  {author} {\bibinfo {author} {\bibfnamefont {E.}~\bibnamefont {Lake}}, \bibinfo {author} {\bibfnamefont {M.}~\bibnamefont {Hermele}}, \ and\ \bibinfo {author} {\bibfnamefont {T.}~\bibnamefont {Senthil}},\ }\bibfield  {title} {Dipolar Bose-Hubbard model,\ }\href {\doibase/10.1103/PhysRevB.106.064511} {\bibfield  {journal} {\bibinfo  {journal} {Phys. Rev. B}\ }\textbf {\bibinfo {volume} {106}},\ \bibinfo {pages} {064511} (\bibinfo {year} {2022})}\BibitemShut {NoStop}%
\bibitem [{\citenamefont {McDonald}\ \emph {et~al.}(2018)\citenamefont {McDonald}, \citenamefont {Pereg-Barnea},\ and\ \citenamefont {Clerk}}]{McDonald2018}%
  \BibitemOpen
  \bibfield  {author} {\bibinfo {author} {\bibfnamefont {A.}~\bibnamefont {McDonald}}, \bibinfo {author} {\bibfnamefont {T.}~\bibnamefont {Pereg-Barnea}}, \ and\ \bibinfo {author} {\bibfnamefont {A.}~\bibnamefont {Clerk}},\ }\bibfield  {title} {Phase-Dependent Chiral Transport and Effective Non-Hermitian Dynamics in a Bosonic Kitaev-Majorana Chain,\ }\href {http://dx.doi.org/10.1103/PhysRevX.8.041031} {\bibfield  {journal} {\bibinfo  {journal} {Physical Review X}\ }\textbf {\bibinfo {volume} {8}} (\bibinfo {year} {2018})}\BibitemShut {NoStop}%
\bibitem [{\citenamefont {Kitaev}(2001)}]{Kitaev2001}%
  \BibitemOpen
  \bibfield  {author} {\bibinfo {author} {\bibfnamefont {A.~Y.}\ \bibnamefont {Kitaev}},\ }\bibfield  {title} {Unpaired Majorana fermions in quantum wires,\ }\href {http://dx.doi.org/10.1070/1063-7869/44/10S/S29} {\bibfield  {journal} {\bibinfo  {journal} {Physics-Uspekhi}\ }\textbf {\bibinfo {volume} {44}},\ \bibinfo {pages} {131–136} (\bibinfo {year} {2001})}\BibitemShut {NoStop}%
\bibitem [{\citenamefont {Li}\ \emph {et~al.}(2022)\citenamefont {Li}, \citenamefont {Cao}, \citenamefont {Chen},\ and\ \citenamefont {Yang}}]{Li2022}%
  \BibitemOpen
  \bibfield  {author} {\bibinfo {author} {\bibfnamefont {Y.}~\bibnamefont {Li}}, \bibinfo {author} {\bibfnamefont {Y.}~\bibnamefont {Cao}}, \bibinfo {author} {\bibfnamefont {Y.}~\bibnamefont {Chen}}, \ and\ \bibinfo {author} {\bibfnamefont {X.}~\bibnamefont {Yang}},\ }\bibfield  {title} {Universal characteristics of one-dimensional non-Hermitian superconductors,\ }\href {http://dx.doi.org/10.1088/1361-648X/aca4b4} {\bibfield  {journal} {\bibinfo  {journal} {Journal of Physics: Condensed Matter}\ }\textbf {\bibinfo {volume} {35}},\ \bibinfo {pages} {055401} (\bibinfo {year} {2022})}\BibitemShut {NoStop}%
\bibitem [{\citenamefont {Yan}\ \emph {et~al.}(2023)\citenamefont {Yan}, \citenamefont {Cui}, \citenamefont {Liu}, \citenamefont {Cao}, \citenamefont {Zhang},\ and\ \citenamefont {Wang}}]{Yan2023}%
  \BibitemOpen
  \bibfield  {author} {\bibinfo {author} {\bibfnamefont {Y.}~\bibnamefont {Yan}}, \bibinfo {author} {\bibfnamefont {W.-X.}\ \bibnamefont {Cui}}, \bibinfo {author} {\bibfnamefont {S.}~\bibnamefont {Liu}}, \bibinfo {author} {\bibfnamefont {J.}~\bibnamefont {Cao}}, \bibinfo {author} {\bibfnamefont {S.}~\bibnamefont {Zhang}}, \ and\ \bibinfo {author} {\bibfnamefont {H.-F.}\ \bibnamefont {Wang}},\ }\bibfield  {title} {Topological phase in a nonreciprocal Kitaev chain,\ }\href {http://dx.doi.org/10.1088/1367-2630/ad1140} {\bibfield  {journal} {\bibinfo  {journal} {New Journal of Physics}\ }\textbf {\bibinfo {volume} {25}},\ \bibinfo {pages} {123023} (\bibinfo {year} {2023})}\BibitemShut {NoStop}%
\bibitem [{\citenamefont {Ardonne}\ and\ \citenamefont {Kurasov}(2025)}]{Ardonne2025}%
  \BibitemOpen
  \bibfield  {author} {\bibinfo {author} {\bibfnamefont {E.}~\bibnamefont {Ardonne}}\ and\ \bibinfo {author} {\bibfnamefont {V.}~\bibnamefont {Kurasov}},\ }\bibfield  {title} {On the non-hermitian Kitaev chain,\ }\href {http://dx.doi.org/10.1088/1751-8121/add4d3} {\bibfield  {journal} {\bibinfo  {journal} {Journal of Physics A: Mathematical and Theoretical}\ }\textbf {\bibinfo {volume} {58}},\ \bibinfo {pages} {205302} (\bibinfo {year} {2025})}\BibitemShut {NoStop}%
\end{thebibliography}%

\newpage

{\bf \center Supplementary material}
\setcounter{equation}{0}
\setcounter{figure}{0}
\renewcommand{\theequation}{S.\arabic{equation}}
\renewcommand{\thefigure}{S.\arabic{figure}}

\begin{figure}[htbp]
    \centering
    \includegraphics[width=1\columnwidth]{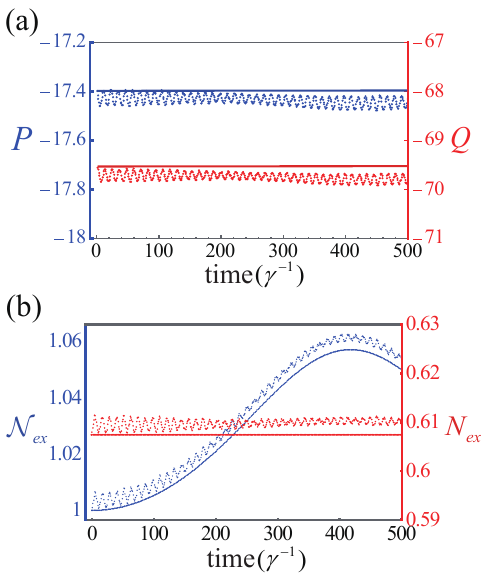}
    \caption{{\bf } {Numerical results of a $7\times 2$ full ladder model $\hat{H}_s+\hat{V}_{on}$ and the comparisons with the results of the effective model $\hat{H}_{ex}$. We choose $\gamma=1$, $\Omega=50, V_0=100, t=1,$ and $\phi=1$. (a) The expectation values of the total dipole moment $P$ and quadrupole moment $Q$ as a function of time. (b) The expectation value of the total exciton density $\sum_j \hat{d}_j^\dagger \hat{d}_j$ and a conserved quantity $\sum_j e^{2A_{ex}j} \hat{d}_j^\dagger \hat{d}_j$ as a function of time. The smooth curves are from the effective model in Eq.~(\ref{Heff}).}}
    \label{PQ}
\end{figure}

In this supplementary material, we discuss conserved quantities in our system by taking into account the finite curvatures underlying Hatano-Nelson chains, and present related results from numerical simulations. We first consider the time evolution of multipole moments. By %treating atoms as charges in classical electrodynamics, 
setting $x=j$ for atoms on the $j$-th site, and $y=\pm1$ for the excited and ground band,  multipole moments may be defined as, %as our observables,
\begin{equation}
    \begin{cases}
        \mathcal{P}=\mathcal{P}_y=\sum_j \hat{n}_{j,+}-\hat{n}_{j,-}\\
        \mathcal{Q}=\mathcal{Q}_{xy}=\sum_j j(\hat{n}_{j,+}-\hat{n}_{j,-}),
    \end{cases}
\end{equation}where $\mathcal{P}$ is the total dipole moment along the y-direction and $\mathcal{Q}$ is the off-diagonal element of the quadrupole moment tensor. However, such definitions of multipole moments in non-Hermitian systems %have a subtle issue, 
face a fundamental issue, as the total charge $\mathcal{N}=\sum_j \hat{n}_{j,+}+\sum_j \hat{n}_{j,-}$ is not conserved. To resolve this issue, we apply a similarity transformation on the fermionic operators in Eq.~(\ref{Ke})
\begin{equation}
    \begin{aligned}
        \hat{c}_{j,s}^\dagger \rightarrow& (t_{L,s}/t_{R,s})^{j/2}\hat{a}_{j,s}^\dagger\\
        \hat{c}_{j,s}\rightarrow& (t_{L,s}/t_{R,s})^{-j/2}\hat{a}_{j,s},\label{similarity}
    \end{aligned}
\end{equation}where $\hat{a}_j$ and $\hat{a}_j^\dagger$ are a new set of fermionic operators that satisfy $\{\hat{a}_j,\hat{a}_m^\dagger\}=\delta_{m,j}$. The effective Hamiltonian in Eq.~(\ref{Ke}) becomes Hermitian,
\begin{equation}
    \hat{K}_e=\sum_{j,s=\pm} -\sqrt{t_{R,s}t_{L,s}}(\hat{a}_{j+1,s}^\dagger \hat{a}_{j,s}+h.c.)+\Omega(\hat{n}_+-\hat{n}_-).\label{KeHermitian}
\end{equation}

As a result, when we describe a quantum state $|\psi\rangle$ in terms of the Fock states generated by $\hat{a}_{j,s}^\dagger$, the total particle number $\langle\psi|\sum_j \hat{a}_{j,s}^\dagger \hat{a}_{j,s}|\psi\rangle$ on each band has to be conserved. In contrast, when $|\psi\rangle$ is expressed in terms of the Fock states generated by $\hat{c}_{j,s}^\dagger$, Eq.~(\ref{similarity}) tells us that an additional normalization factor $(t_{L,s}/t_{R,s})^{j}$ has to be considered when we calculate the expectation values of $\sum_j \hat{c}_{j,s}^\dagger \hat{c}_{j,s}$. %For convenience, we suppose all the Fock states have already been normalized and 
We extract the normalization factor of each Fock state as an operator $\hat{\eta}_s=(t_{L,s}/t_{R,s})^{\sum_m m \hat{c}_{m,s}^\dagger \hat{c}_{m,s}}$. The conserved total particle number is thus expressed as $\langle \psi |\hat{\eta}_s\sum_j \hat{c}_{j,s}^\dagger \hat{c}_{j,s}|\psi\rangle $, where $\hat{\eta}_s$ serves as a metric operator that modulates the density distribution of many-body quantum states. The operator $\hat{\eta}_s$ can be understood from the finite curvature underlying the non-Hermitian model~\cite{Lv2022}.

Correspondingly, we define the multipole moments as
\begin{equation}
    \begin{cases}
        P=\langle \psi|\sum_j (\hat{\eta}_+\hat{n}_{j,+}-\hat{\eta}_-\hat{n}_{j,-})|\psi\rangle\\
        Q=\langle \psi|\sum_j j(\hat{\eta}_+\hat{n}_{j,+}-\hat{\eta}_-\hat{n}_{j,-})|\psi\rangle.
    \end{cases}
\end{equation}In the numerical simulation, 
we computed the time evolution of the initial state in Eq.~(\ref{init}) for both the effective model in Eq.~(\ref{Heff}) and the full model $\hat{H}_s+\hat{V}_{on}$. The results of $P$ and $Q$ are shown in Fig.~\ref{PQ}a. As expected, the results of the effective model are time-independent constants, confirming the above analysis. If we consider the full model, the results %We first show the expectation value of total dipole moment $P$ (blue) and quadrupole moment $Q$ (red) as a function of time in Fig.~\ref{PQ}a. They are approximately conserved quantities with 
show 
small oscillations about the mean results. Such oscillations originate from the corrections to the effective model when $\Omega$ is finite. 
%originating from the Rabi oscillation between the $+$ and $-$ bands when $\Omega$ is finite. 
With increasing $\Omega\rightarrow\infty$, %the bands become decoupled and the 
the effective model works better and the small
oscillations are suppressed down to zero, recovering the results of the effective theory. %. This observation is consistent with our expectation since the effective Hamiltonian in Eq.~(\ref{Ke}) and (\ref{KeHermitian}) are based on the perturbation theory in the large $\Omega$ limit. Eq.~(\ref{KeHermitian}) indicates that the tunneling of atoms is reciprocal in a curved space, such that the dipole moment and quadrupole moment should be conserved. 

From the perspective of a single exciton, %dynamics, 
the metric operator $\hat{\eta}$ reduces to $e^{2A_{ex}j}$ where $A_{ex}$ is the imaginary vector potential seen by the exciton and $j$ labels the position of the exciton. If such $\hat{\eta}$ is not included in the definition of the exciton number, 
\begin{equation}
    \mathcal{N}_{ex}=\langle \psi| \sum_j \hat{d}_j^\dagger \hat{d}_j |\psi\rangle. 
\end{equation}
Blue data in Fig.~\ref{PQ}b shows that $\mathcal{N}_{ex}$ is not conserved. Including $\hat{\eta}$, we obtain, 
\begin{equation}
    {N}_{ex}=\langle \psi| \sum_j e^{2A_{ex}j}\hat{d}_j^\dagger \hat{d}_j |\psi\rangle, 
\end{equation}
and red data in Fig.~\ref{PQ}b shows that ${N}_{ex}$ becomes time-independent. Again, the small difference between the results of the full model and the effective model comes from the corrections to the effective model when $\Omega$ is finite. With increasing $\Omega$, the small oscillations in the results of the full model vanish and the results of the full model reduce to those of the effective model.  

%In Fig.~\ref{PQ}b, we show the time evolution of the exciton number. Without $\hat{\eta}$ (blue),  

%The data with oscillations are results of the full model in Eq.~(\ref{HL}), while the smooth curves are results of the effective model in Eq.~(\ref{Heff}). The effective model captures well the dynamics of the total exciton density in our parameter regime. After taking into account the metric operator, the total exciton density is indeed conserved.

\end{document}